\documentclass[aps,prd,preprint,nofootinbib,citeautoscript,floatfix]{revtex4-1}

\usepackage[export]{adjustbox}
\usepackage{amsmath,amsfonts,amssymb,amsthm,empheq}
\usepackage{color}
\usepackage{graphicx}
\usepackage{dcolumn}
\usepackage{bm}
\usepackage{hyperref}
\usepackage{natbib}
\usepackage{float}
\usepackage[caption=false]{subfig}
\usepackage{hyperref}
\hypersetup{
    colorlinks=true,
    linktoc=page,    
    linkcolor=blue,
    urlcolor=magenta
}

\newcommand{\be}{\begin{equation}}
\newcommand{\ee}{\end{equation}}
\newcommand{\bea}{\begin{eqnarray}}
\newcommand{\eea}{\end{eqnarray}}
\newcommand{\mM}{\mathcal{M}}

\newcommand{\mD}{\mathcal{D}}

\newcommand{\pd}{\partial}
\newcommand{\nn}{\nonumber}

\newcommand{\mS}{\mathcal{S}}

\newcommand{\A}{\text{Area}}

\newcommand{\newe}{\nn\\&=&}

\begin{document}

\title{Gauge theories with nontrivial boundary conditions: \\Black holes}
\author{Peng Cheng}
\email{p.cheng.nl@outlook.com}
\affiliation{
Center for Joint Quantum Studies and Department of Physics, School of Science, Tianjin University, 135 Yaguan Road, Tianjin 300350, China}
\affiliation{
     Institute for Theoretical Physics, University of Amsterdam, 1090 GL Amsterdam, Netherlands}
\begin{abstract}

We study the partition function and entropy of U(1) gauge theories with multiple boundaries on the black holes background. The nontrivial boundary conditions allow residual zero longitudinal momentum modes and Wilson lines stretched between boundaries. Topological modes of the Wilson lines and other modes are also analyzed in this paper. We study the behavior of the partition function of the theory in different temperature limits, and find the transitions of dominances of different modes as we vary the temperature. Moreover, we find two different area contributions plus logarithm corrections in the entropy. One being part of the bulk fluctuation modes can be seen for finite-temperature black holes, and the other coming from vacuum degeneracy can only be seen in the superlow temperature limit. 
We have confirmed the mechanism and entropy found in the superlow temperature limit also persist for extremal black holes.
The gauge fluctuation on the black hole background might help us understand some fundamental aspects of quantum gravity related to gauge symmetries.
\end{abstract}

\maketitle
\flushbottom

\newpage
\section{Introduction}
\label{intro}

It is well known that black holes are thermodynamic systems as seen by outside observers \cite{Hawking1971,Christodoulou1970,Christodoulou1971,HAWKING1974,Bekenstein1973,Bardeen1973,Bekenstein1972,CARTER1972}, which might reflect the properties of the underlying microscopic structures. The so-called central dogma \cite{Almheiri2020} claims that the number of the microscopic degrees of freedom of a black hole should be the Bekenstein-Hawking entropy
\be
S=\frac{\text{Area}}{4 \hbar G_N}\propto \frac{\text{Area}}{l_p^2}\,,\label{Bekentropy}
\ee
and the whole system unitarily evolves under time evolution. $l_p$ is the Planck length.

So, one of the most important questions in black hole physics is ``what is the theory that describes the microscopic structure of a black hole".
Many different research programs are trying to give a microscopic explanation and reproduce the Bekenstein-Hawking entropy by counting the microstates. 
Those research programs can be more or less divided into two categories.
\begin{itemize}
  \item The first train of thought is to add or find microscopic states near the horizon. Usually, one expects the quantum fields living close to the horizon or entanglement pairs across the horizon would have an entropy contribution proportional to the area \cite{tHooft:1984kcu, Srednicki1993, Susskind1994, Solodukhin2011, Blommaert2018a, Donnelly:2014fua, Donnelly2015a, Wei2015am, Haco2018}. 
	There are more proposals, like loop quantum gravity \cite{Krasnov1996, Rovelli1996, Ashtekar1997, Domagala2004, Meissner2004, Ghosh2006} that can also be categorized here.
  \item The second set of ideas is to explain the Bekenstein-Hawking entropy by finding hidden symmetries, which are mainly used to understand the entropy of (near-)extremal black holes.
  To reproduce the universal coefficient $1/4$ in the Bekentein-Hawking entropy, one always needs to use a powerful tool: symmetry. If there are hidden conformal symmetries near the horizon, the density of states of two-dimensional CFT is controlled by the Cardy formula \cite{Cardy1986, Bloete1986}. The universality of the Cardy formula can be used to reproduce the Bekenstein-Hawking entropy with the coefficient $1/4$ \cite{Strominger1998, Birmingham1998, Guica2009, Carlip2011}. 
Hidden symmetries (can also be Bondi-Metzner-Sachs, Carrollian, and others) and their relation with black hole entropy are extensively studied in the literature \cite{Brown1986, Solodukhin1999, Carlip1999a, Carlip1999, Silva2002, Carlip2007, Castro2010, Wang2010, Chen2010, Carlip2011a, Compere2017, Song2012, Carlip2018, Carlip2019}. 
  The holographic or stringy methods \cite{Strominger1996, Callan1996, Peet2000, Das2001, Maldacena1996, Emparan2006} also heavily rely on the symmetries of spacetime.
However, those methods related to symmetries usually work only in (near-)extremal black hole cases, and are difficult to be generalized to finite-temperature black holes.
\end{itemize}

Despite completely different starting points, lots of theories in both categories successfully reproduced the Bekenstein-Hawking entropy.
Is it a problem to have so many different explanations for black hole microstates?
There are two obvious responses to the above question. i) There are common structures that lie behind those theories, and we may find a unified description that explains why those different theories are able to count the microscopic structures of black holes. ii) Some of the theories describe the wrong physics, and the corresponding entropy was reproduced by counting fake microscopic structures of the black hole.
This paper aims to address this problem using a toy model of gravitational fluctuations on a black hole background. Our final conclusion tends to say that the two categories of theories are both counting the right microscopic structures of the black hole although they are counting different things. Moreover, the microstates that are used to explain the entropy of finite temperature black holes are not the same ones for extremal black holes.

Gauge theories on a fixed background can be regarded as good toy models of gravitational fluctuations around saddle points of gravity theory.
The one-loop correction around a fixed saddle is captured by the linearized Einstein-Hilbert action, which is a massless quadratic Fierz-Pauli action.
The massless Fierz-Pauli theory is a gauge theory with two physical polarizations; thus, we can use gauge theories on classical solutions of Einstein's theory as toy models of metric perturbations.
The U(1) gauge theory has a simpler structure to deal with and can properly capture the gauge subtleties of the gravity theory, which is the theory we will focus on in this paper.

As a brief summary of the main results, we study the Euclidean path integral of U(1) gauge theory living between two parallel boundaries. The system is put on a black hole background with the boundaries perpendicular to the radius direction of the black hole $r$. We allow residual degrees of freedom of $A_r$ on the boundaries.  
In the first paper of the series \cite{Cheng2022}, we studied the flat case with two boundaries and found nonlocal effects due to the interplay between the boundaries. Here, the analysis is generalized to the curved spacetime, and we carefully study different behavior of the gauge theory on the black hole background.
The physical variables of the theory are bulk fluctuation modes, zero longitudinal momentum mods of $A_r$, Wilson lines stretched between the two boundaries, and other modes.\footnote{Like the topological modes of the Wilson lines.}

The bulk fluctuation modes whose entropy scales as the volume of the region are always dominant at high temperatures. The bulk part also contains a contribution proportional to the area of the horizon plus logarithm corrections due to the highly curved spacetime near the horizon. As the temperature cools down, we would mainly see the fluctuation modes of edge residual degrees of freedom and the Wilson lines. The entropy of those modes scales as the area of the boundaries multiplied by the temperature squared. For superlow temperatures, there is a localization of modes on $\pd_\tau \phi=\pd_\tau W=0$. Performing the path integral, the corresponding entropy of those modes scales as the horizon area divided by the Planck length squared. The constant modes and topological modes of the boundary-stretched Wilson lines contribute logarithm corrections to the entropy at superlow temperatures. So, we also get area contribution plus logarithm corrections at superlow temperatures. 
 
Now, we have seen that the Bekenstein-Hawking-like entropies and logarithm corrections come from two different places.
 The first contribution comes from the bulk fluctuation modes, which agrees with the entropy found in the ``brick wall" model \cite{tHooft:1984kcu}. Those degrees of freedom correspond to the degrees of freedom living very close to the horizon.
 The second contribution is because the zero longitudinal momentum modes of $A_r$ and the boundary-stretched Wilson lines $W$ are localized in the phase space at $\pd_\tau \phi=\pd_\tau W=0$, whose entropy is coming from the zero-point energy. Those modes correspond to the vacuum degeneracy near the horizon and can be explained by symmetry-breaking patterns. 
Note that the fluctuation modes dying off at superlow temperature was well-studied \cite{Amsel:2009ev, Dias:2009ex,Johnstone:2013ioa, Hajian:2013lna, Hajian:2014twa}, so the extremal black hole entropy coming from a different mechanism, like the zero-point energy discussed in the current paper, is a reasonable way out.

It is interesting to notice that the first contribution only appears in finite-temperature black holes, and the second one only appears in the superlow temperature limit.
We thus infer that the two categories of theories might correspond to different modes found in this calculation.
The attempts that are trying to add or find quantum structure near the horizon are counting the same thing as the contributions contained in the bulk fluctuation modes. 
Furthermore, the attempts that are trying to find hidden symmetries for extremal black holes and explaining the black hole entropy from a symmetry-breaking viewpoint are counting the same thing as the modes dominating at the superlow temperatures. 
The black hole entropy for finite-temperature black holes and extremal black holes might be counting completely different things.

In this paper, we study the partition function of U(1) gauge theories with multiple boundaries on a black hole background. The paper is organized as follows. 
In Sec. \ref{review}, we briefly review the main results from the first paper of this series, where we studied the flat case. The flat case can be regarded as a guide for our black hole calculation.
And then, we analyze the allowed boundary conditions and derive effective actions for different modes in Sec. \ref{finite}. We perform the Euclidean path integral in Sec. \ref{behavior}, and discuss different behavior of the fields at different temperatures.
The same mechanism is confirmed in the extremal black hole case in Sec. \ref{extremal}.
Sec. \ref{conclusions} is a summary of the whole paper and we also provide some further physical discussion there.
Appendix \ref{solB} is devoted to studying a different but still interesting set of boundary conditions. More details of deriving the effective action are exiled to the Appendixes \ref{appA2} and \ref{redim}. The path integral of fluctuation modes in curved spacetime is demonstrated in Appendix \ref{Sch}.

\section{A brief review of the flat case}
\label{review}

\begin{figure}[H]
  \centering
  \includegraphics[width=0.6\textwidth]{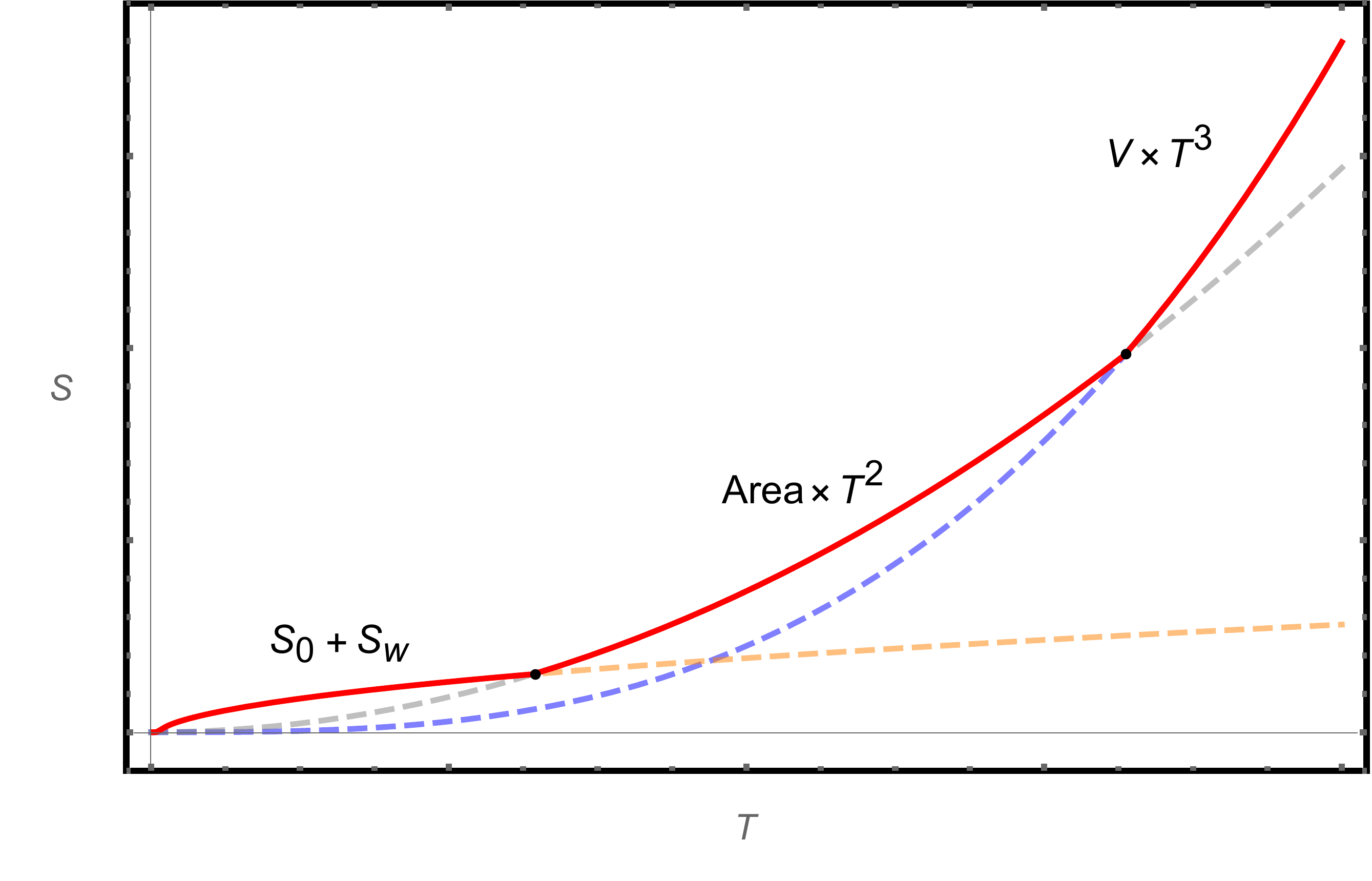}\\
  \caption{Sketch of the entropy of gauge fields with different temperatures in the flat case. The red line demonstrates the dominant contribution, and the overall entropy is the sum of different contributions.}\label{boxphase}
\end{figure}

In the first paper of the series \cite{Cheng2022}, we studied gauge theories between two parallel boundaries with nontrivial boundary conditions on a flat background. We allowed residual degrees of freedom of $A_r$ to survive on the boundary, with the direction perpendicular to the boundary denoted as $r$. Besides the residual modes due to the boundary conditions, we also found Wilson lines stretched between different boundaries because the interplay between boundaries is an interesting mode of the system. The symplectic form and the canonical commutation relations were carefully studied, which helped us to confirm the dynamical variables and measures in the phase space. It was shown that there are bulk fluctuation modes, boundary edge modes, Wilson lines, and other interesting modes in the phase space. With the canonical analysis, we derived the partition function and entropy via the Euclidean path integral method, and studied the different behavior of those modes by varying the temperature of the system.

 The main results are shown in Fig. \ref{boxphase}. 
 At high temperatures, the partition function and entropy are the standard ones for blackbody radiation at temperature $T$. The entropy contains two copies of physical polarizations and is proportional to the volume between the two boundaries multiplied by $T^3$. 
 As the temperature becomes cooler than before, the surviving zero longitudinal momentum modes $\phi$ start to be the most important contribution. 
 Also, boundary-stretched Wilson lines $W$ have similar behavior. $\phi$ and $W$ just behave like two lower-dimensional massless scalar fields living on the boundary. 
 At relatively high temperatures, the fluctuation modes of $\phi$ and $W$ are dominant and their entropies scale as the area of the boundary multiplied by $T^2$. 
 At superlower temperatures, all the fluctuation contributions die off; and we are left with the constant and winding modes of $W$, whose entropies more or less scale as the logarithm of the length scales of the theory, as shown in Fig. \ref{boxphase}. Note that the winding modes of $W$ arise due to the map between the Euclidean time circle and U(1) symmetry.
 
 There are two transitions of dominance as can be seen from Fig. \ref{boxphase}. The bulk fluctuation modes always dominate in the high-temperature limit, and the entropy scales as the volume multiplied by temperature cubed. As lower temperature comes, the area contribution starts to dominate. At superlow temperatures, the fluctuation contribution is not important anymore, and the only contribution is from the constant modes and winding modes of the field $W$.
 
 The flat case was regarded as a toy model of the more general situation, for example, curved spacetime as the background. With the canonical analysis and partition function calculation of the flat case, we expect to see a similar set of dynamical variables in the phase space. Moreover, we can follow a similar logic in decomposing the variables and calculating the partition function and entropy.

\section{Gauge fields on Schwarzschild black hole}
\label{finite}

Black holes are systems associated with temperatures and entropies, and the Euclidean method for finite temperature systems also works for black holes. 
However, the black hole system is special compared to the flat case discussed, because the Bekenstein-Hawking entropy is always proportional to the area of the horizon in units of $G_N$.
As has been reviewed in the previous section, there are interesting transitions of dominances of different modes arising due to different reasons,
one might speculate that a similar phenomenon would also happen in black hole systems, which might explain something deep in the black hole mechanism.

\begin{figure}
  \centering
  \includegraphics[width=0.5\textwidth]{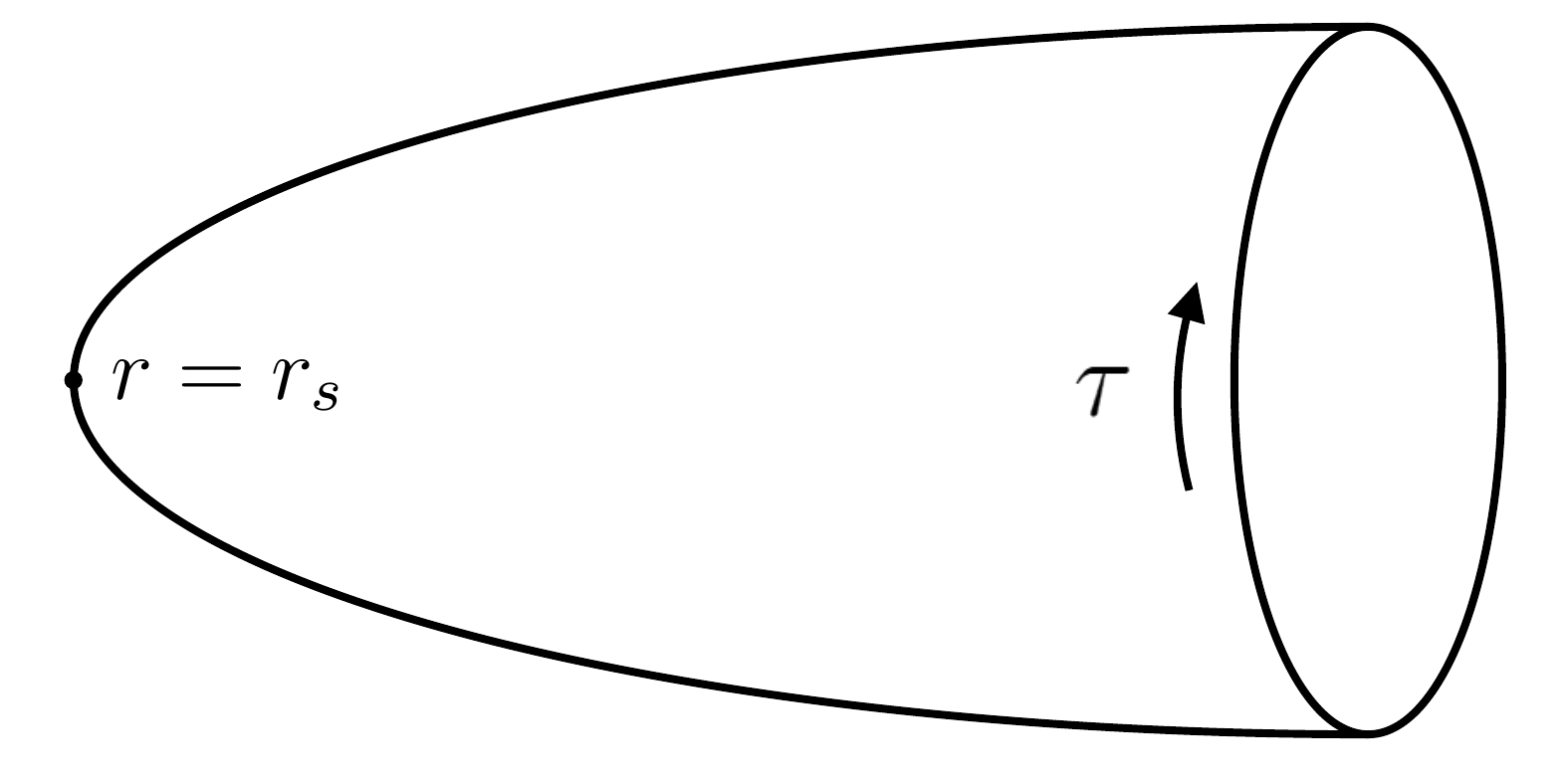}\\
  \caption{Geometry of a Euclidean black hole, where every point on the cigar is an ${S}^2$. We put two boundaries on this background, at the stretched horizon ($r=r_s+\varepsilon$) and distance $L$ away from the horizon ($r=r_s+L$) separately.}\label{cigar}
\end{figure}

\begin{figure}
  \centering
  \includegraphics[width=0.4\textwidth]{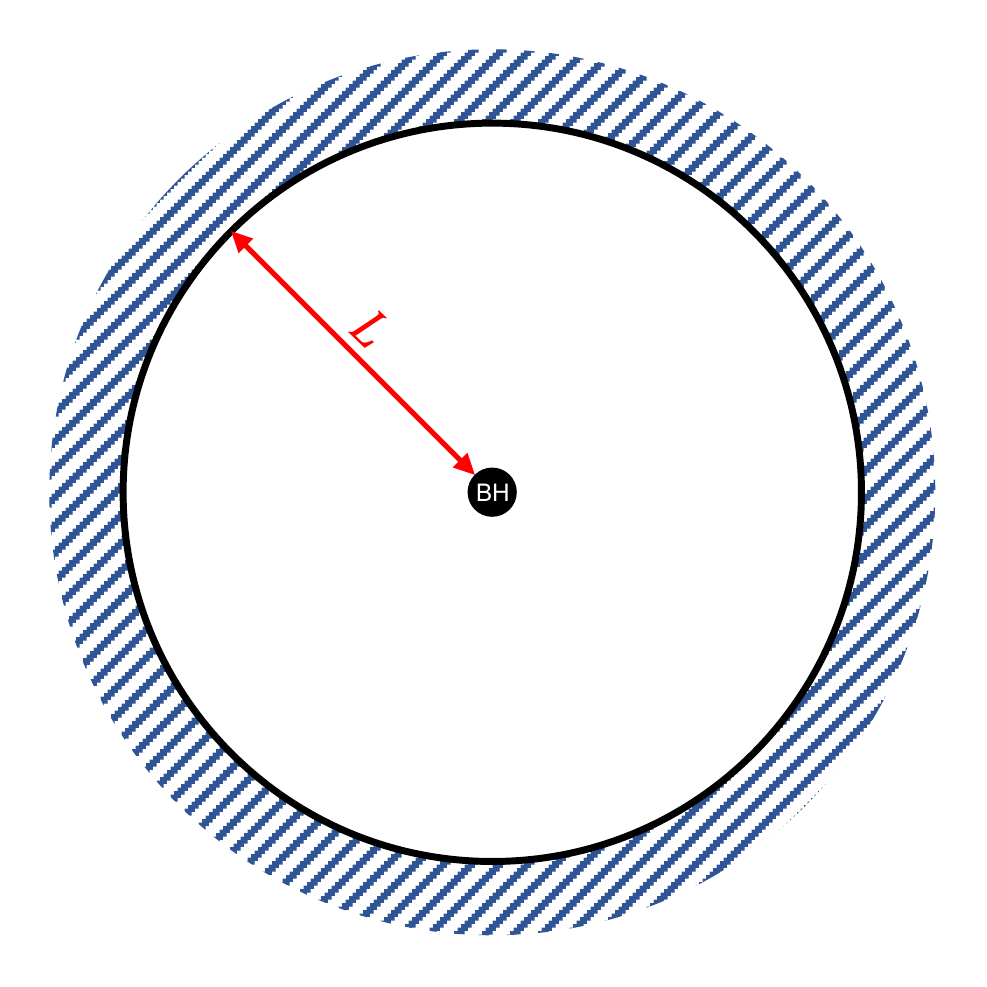}~~~
  \includegraphics[width=0.4\textwidth]{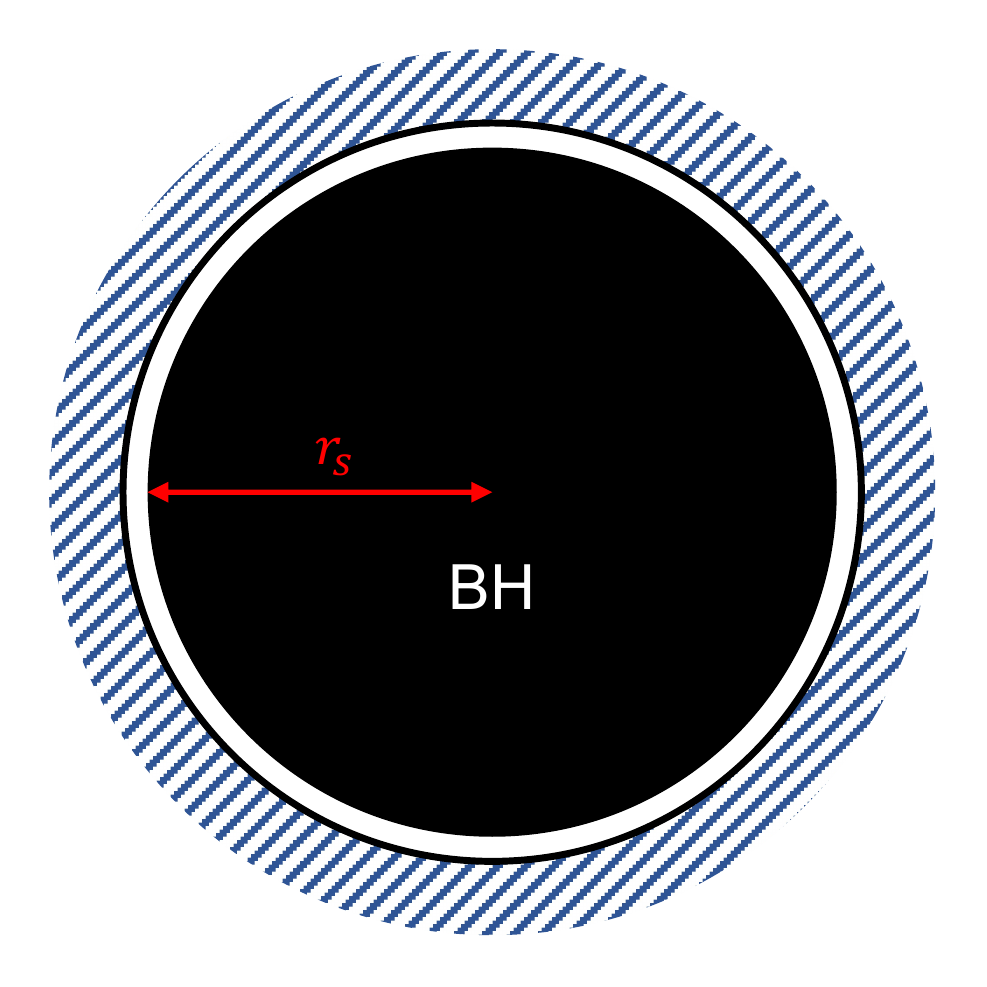}\\
  \caption{Different temperature limits of black holes in a box. \textbf{Left}: high temperature limit $L\gg r_s$, i.e. the small black hole case. The temperature of the system is $T\propto 1/r_s$. The mere role of the black hole is to give a very high temperature $T$ to the system. We expect the final results to be similar to the high-temperature limit of the flat case. \textbf{Right}: low temperature limit of the system $L\ll r_s$, i.e. large black hole case. We expect the fluctuation modes should be less important, but we still see Bekenstein-Hawking-like entropy.}\label{highlowT}
\end{figure}

The background we are interested in is a Euclidean Schwarzschild black hole with metric
\be\label{metric}
ds^2=(1-\frac{r_s}{r})d\tau^2+ (1-\frac{r_s}{r})^{-1}dr^2 +r^2 d\Omega^2\,,
\ee
where the Schwarzschild radius is $r_s=2G_N M$. 
Note that one important difference between the flat case and black hole case is that we are using spherical coordinates rather than Cartesian here.
The inverse temperature of the system \eqref{metric} is identified with the periodicity of the Euclidean time $\beta$, to avoid conical singularity at the horizon.
The geometry can be illustrated as the so-called ``cigar" geometry, as shown in Fig. \ref{cigar}. 
Following 't Hooft \cite{tHooft:1984kcu}, we put two boundaries on this background and study different temperature limits of the system, as shown in Fig. \ref{highlowT}.
The first boundary is located at a very small distance $\varepsilon$ away from the horizon, known as the ``stretched horizon" \cite{Susskind:1993if}. 
The other boundary is at the distance $L$ away from the horizon, which is the surface of the box similar to the flat case. 

We will mainly discuss two different limits of the black hole system, as shown in Fig. \ref{highlowT}.
In the high-temperature limit $L\gg r_s $, shown in the first panel, the situation is supposed to be similar to the flat case because the existence of the small black hole merely gives a periodicity to the Euclidean time, i.e a temperature $T=1/\beta$, to the system. 
On the other hand, in the low temperature limit $L\ll r_s$, we expect the results to be qualitatively different from the high-temperature limit and share some similarities with the extremal black hole case, the behavior of which was also well studied in literature \cite{Strominger1996, Callan1996, Maldacena:1996ds, Maldacena:1997ih, Peet2000, Banerjee:2010qc, Sen:2012cj, Iliesiu:2020qvm}. 
This means that there will be transitions of different behaviors, which might be explained by the symmetry-breaking pattern.

We can always redefine $\rho=r-r_s$ and write the metric as
\be
\label{metricrho}
ds^2=\frac{\rho}{\rho+r_s}d\tau^2+ \frac{\rho+r_s}{\rho}d\rho^2 +(\rho+r_s)^2 d\Omega^2\,.
\ee
Now the two boundaries are at $\rho=\varepsilon$ and $\rho=L$ separately. 
In order to derive the lower-dimensional effective action by dimensional reduction, it is convenient to define the proper distance $y$ along the radius direction
\be
dy=\sqrt{\frac{\rho+r_s}{\rho}}d\rho\,.
\ee
The proper distance $y$ can be integrated out and read as
\be
y=\sqrt{\rho~(\rho+r_s)}+r_s\text{~arcsinh} \sqrt{\frac{\rho}{r_s}}\,.
\ee
Inverting the above equation, one can express $\rho$ as a function of $y$ and rewrite the black hole metric as
\be\label{yymetric}
ds^2=\frac{\rho(y)}{\rho(y)+r_s}d\tau^2+ dy^2 +[\rho(y)+r_s]^2 d\Omega^2\,,
\ee
which will be the metric we mainly work with. Note that $y$ takes value from $y_1$ to $y_2$, with
\be
y_1=\sqrt{\varepsilon(\varepsilon+r_s)}+r_s\text{~arcsinh} \sqrt{\frac{\varepsilon}{r_s}}\,,~~~~~~y_2=\sqrt{L(L+r_s)}+r_s\text{~arcsinh} \sqrt{\frac{L}{r_s}}\,.
\ee

Before moving to the gauge theory and boundary conditions, let us briefly discuss issues related to the stretched horizon.
Here the stretched horizon can be regarded as an auxiliary timelike surface, where we can put boundary conditions on. 
There are different opinions towards the stretched horizon
\footnote{Mathematically, one can put boundaries at any location. The original physics can be recovered by gluing the two sides along the edge, which means one needs to integrate over the boundary conditions along the boundary. So it is not strange to put a boundary near the horizon. However, whether is it reasonable to put a physical boundary near the horizon is unknown because any object near the horizon would fall into the black hole. 
Although we largely mimic the flat case calculation, one may need to keep in mind that the black hole situation might not be the same as the case of the parallel plates.}; for example, it is believed that only the physics outside of the stretched horizon is well described by local QFT. In this case, the stretched horizon is a physical surface \cite{Susskind2004}, and $\varepsilon$ can be regarded as the UV regulator of the QFT outside of the horizon and can be taken to but not exceed the Planck scale $l_p$.
Whether the stretched horizon is a physical surface or an auxiliary surface does not make much difference for our calculations here, and $\varepsilon$ will be made very small either way.
However, one does need to pay attention to the order of taking different limits of length scales.
As we will see, in the low-temperature limit $\beta\gg L$, taking the $\varepsilon \to 0$ limit would introduce localizations on the spaces of some specific modes, which is essential to get the area entropy contribution. 
However, one may not see this phenomenon if the order of taking the above limits is messed up.

\subsection{Boundary conditions}

Now, let us put U(1) gauge theories between the boundaries, with Euclidean action
\be\label{action2}
S_E=\frac{1}{4e^2}\int_{\mathcal{M}}d^4x \sqrt{g} ~F^{\mu\nu}F_{\mu\nu}\,.
\ee
As analyzed in \cite{Cheng2022}, to have a well-defined variation principle, there are two sets of interesting boundary conditions.
\begin{itemize}
  \item The most interesting boundary condition for this paper is the situation where we allow residual $A_r$ modes on boundaries
  \be\label{first}
  \delta A_{a}\Big{|}_{\pd \mM}=0,~~~~~~\delta A_r\Big{|}_{\pd \mM}=f(x^a)\,,
  \ee
  where $f(x^a)$ can have local dependence of $x^a=(\tau,\theta,\phi)$ and $A_{a}\big{|}_{\pd \mM}$ are fixed configurations. 
  \item The other interesting boundary condition is Neumann-like:
  \be\label{Neumann}
   F^{\mu\nu}\Big{|}_{\pd \mM}=0\,.
  \ee
  There can be interesting physics as well, for example, the would-be pure gauge modes on the boundary are not fixed by the boundary condition. Although this boundary condition is not the main focus of this paper, we will briefly discuss the physics related to \eqref{Neumann} in Appendix \ref{solB}.
\end{itemize}

We are mainly going to focus on the boundary condition \eqref{first} on the black hole background. To be precise, the boundary condition $A_a$ components are fixed to
\bea
A_a\Big{|}_{y=y_{\alpha}} &=& 0\,,\label{BHbdy}
\eea
where $\alpha$ labels different boundaries. As for the $A_y$ component, we can separate it as
\be\label{decom}
A_y(x^{\mu})=\hat{A}_y(x^\mu)+\frac{\phi(x^a)}{|y|}\,,
\ee
with $\hat{A}_y(x^\mu)\big{|}_{y=y_1}=0$ and $|y|=y_2-y_1$. So, $\phi(x^a)$ more or less captures the degrees of freedom at the left boundary $y=y_1$. Besides $\phi(x^a)$, there are residual modes at the boundary $y=y_2$.
Following a similar logic from the flat case, the difference between the two boundaries can be captured by Wilson lines stretched between the two boundaries
\be
W(x^a)=i\int_{y_1}^{y_2}dy ~\hat A_y.
\ee
The boundary-stretched Wilson lines capture the physical non-local effects and will play an important role in our main calculation.
There is no constraint on the boundary configurations of $A_y$, so we need to integrate all the boundary configurations of $A_y$ into the path integral.

One can work out the Hamiltonian formulation and symplectic form to study the phase space of the theory with the indicated boundary condition. The dynamical variables in the phase space would be clear from the canonical analysis, which tells us which degrees of freedom should be included in the path integral formulation. 
We have done an analog analysis for the flat case in paper \cite{Cheng2022}, while we will not go through all the canonical analysis here. 
Similar to the flat case, we would also have bulk modes $\hat A_\mu(x^{\mu})$, zero momentum modes along the $y$ direction $\phi(x^a)$, and Wilson lines stretched between the two boundaries $W(x^a)$ on a black hole background. 
We will justify that they are physical by working out their effective actions.

\subsection{The effective action}
\label{seceff}
The partition function can be calculated by the Euclidean functional integral over all the dynamical variables in the theory. For the theory we are studying, we have bulk fluctuation modes $\hat A_\mu$ and the modes arise because of the nontrivial boundary condition: $\phi$ and $W$.
So the partition function can be written as
\be
Z=\int \mD \hat A_\mu ~\mD \phi ~\mD W~\exp\left\{-S_E[\hat A_\mu, \phi, W]\right\}\,.
\ee
The effective action $S_E$ is worked out in Appendix \ref{appA2}, which reads as
\bea\label{BHSE}
S_{E} &=& \frac{1}{4e^2}\int_\mM d\tau d^3x\sqrt{g}~ \hat F^{\mu\nu}\hat F_{\mu\nu}
+\frac{1}{2e^2 |y|^2}\int_\mM d\tau d^3x \sqrt{g} \left[ g^{ab}\pd_a\phi\pd_b\phi\right]
\,\nn\\&~&
-\frac{1}{e^2|y|}\int d\tau d^2x\left[\sqrt{g} ~ g^{ab}\right]_{y=y_1}^{y=y_2}i\pd_a W \pd_b \phi \,.
\eea
The first part can be directly put into the path integral, so we can write the partition function for the bulk fluctuation modes as
\be
Z_{\hat A}=\int\mD\hat A_\mu\exp\left[-\frac{1}{4e^2}\int_\mM d\tau d^3x\sqrt{g}~ \hat F^{\mu\nu}\hat F_{\mu\nu}\right]\,.
\ee
We will discuss bulk gauge fixing condition and evaluate the above partition function in the next subsection.
The rest of the action is
\bea\label{4daction}
S[\phi,W] = \frac{1}{2e^2 |y|^2}\int_\mM d\tau d^3x \sqrt{g} \left[ g^{ab}\pd_a\phi\pd_b\phi\right]
-\frac{1}{e^2|y|}\int d\tau d^2x\left[\sqrt{g} ~ g^{ab}\right]_{y=y_1}^{y=y_2}i\pd_a W \pd_b \phi\,.
\eea
Note that $S[\phi,W]$ contains three-dimensional and four-dimensional parts, while the fields $\phi(x^a)$ and $W(x^a)$ only depend on the transverse coordinates $x^a$.
In order to perform the path integral, we are going to rewrite $S[\phi,W]$ into a three-dimensional action
\bea
S^{(3)}[\phi,W] &=& \frac{1}{2e'^2}\int d\tau d^2x \sqrt{h} \left[ h^{ab}\pd_a\phi\pd_b\phi\right]-\frac{i}{2e'^2}\int d\tau d^2x \sqrt{h}\label{3daction}\nn\\&~& \times\left[  \gamma_1~ h^{\tau\tau} \pd_\tau W\pd_\tau \phi +\gamma_2~ h^{\theta\theta} \pd_\theta W\pd_\theta \phi+\gamma_3~ h^{\varphi\varphi} \pd_\varphi W\pd_\varphi \phi\right]\,.\label{effec3}
\eea
$e'$, $\gamma_1$, $\gamma_2$, and $\gamma_3$ are undetermined parameters, which will be worked out soon.
This is more or less like the Kaluza-Klein reduction from a higher-dimensional action to a lower-dimensional one.

Now we are going to work out all the parameters in action \eqref{effec3}. More details of the calculations can be found in Appendix \ref{redim}. Before actually doing that, let us first assume that the three-dimensional theory uses the same time coordinate $\tau$ as the original one, and the three-dimensional metric takes the following form
\be
h_{ab}=\text{diag}(h_{\tau\tau},~R^2,~R^2\sin^2\theta)\,.
\ee
$R^2$ is a parameter in the metric that we are going to fix by dimensional reduction. With this assumption, we have $\sqrt{h}=\sqrt{h_{\tau\tau}}~R^2\sin\theta$.
To match the first parts in actions (\ref{3daction}) and (\ref{4daction}), we need to work out the following problem
\bea
\frac{1}{2e^2 |y|^2}\int d\tau d^2x ~(\int_{y_1}^{y_2} dy \sqrt{g}g^{ab}) ~\pd_a\phi\pd_b\phi &=& \frac{1}{2e'^2}\int d\tau d^2x \sqrt{h} \left[ h^{ab}\pd_a\phi\pd_b\phi\right]\,
\eea
The solution can be easily obtained and can be written as
\bea
h_{\tau\tau} = \frac{L R^2}{F}\,,~~~~~~~~~
\frac{1}{e'^2} = \frac{L}{e^2|y|^2}\sqrt{h^{\tau\tau}}=\frac{1}{e^2|y|^2}\frac{\sqrt{L~F}}{R}\,,
\eea
with
\be
F\approx ~3r_s^2L+\frac{3}{2}r_s L^2+\frac{L^3}{3}+r_s^3\ln{L}/{\varepsilon}\,.
\ee
More details can be found in Appendix \ref{redim}.
Now we have obtained the three-dimensional metric $h_{ab}$ and the effective coupling constant $e'^2$. The next step is to calculate the parameters $\gamma_1$, $\gamma_2$, and $\gamma_3$ in the action (\ref{3daction}). We need to match the rest of the action in (\ref{4daction}) and (\ref{3daction}), i.e.
\bea
\frac{1}{e^2 |y|}\int d\tau d^2x\left[\sqrt{g} ~ g^{\tau\tau}\right]_{y=y_1}^{y=y_2}\pd_\tau W \pd_\tau \phi &=& \frac{\gamma_1}{2e'^2}\int d\tau d^2x \sqrt{h} ~\left[  h^{\tau\tau} \pd_\tau W\pd_\tau \phi \right]
\,,\nn\\
\frac{1}{e^2 |y|}\int d\tau d^2x\left[\sqrt{g} ~ g^{\theta\theta}\right]_{y=y_1}^{y=y_2}\pd_\theta W \pd_\theta \phi &=& \frac{\gamma_2}{2e'^2}\int d\tau d^2x \sqrt{h} ~\left[  h^{\theta\theta} \pd_\theta W\pd_\theta\phi \right]\,,\label{g3}\\
\frac{1}{e^2 |y|}\int d\tau d^2x\left[\sqrt{g} ~ g^{\varphi\varphi}\right]_{y=y_1}^{y=y_2}\pd_\varphi W \pd_\varphi \phi &=& \frac{\gamma_3}{2e'^2}\int d\tau d^2x \sqrt{h} ~\left[  h^{\varphi\varphi} \pd_\varphi W\pd_\varphi \phi \right]\,.\nn
\eea
The solution of the equations (\ref{g3}) can be obtained as
\bea
\gamma_1 = -2|y|\sqrt{\frac{r_s}{\varepsilon}}\frac{r_s^2 }{F}\,,~~~~~~~~
\gamma_2 =\gamma_3 = \frac{2|y|}{L}\sqrt{\frac{L}{L+r_s}} \,.
\eea
Now, all the undetermined parameters in the effective action \eqref{effec3} are worked out.

With the effective action \eqref{effec3} at hand, we can perform the path integral to calculate the partition function.
The Gaussian path integral for field $\phi$ can be easily worked out, and we get an effective action for field $W$ in the meantime. The path integral over $\phi$ gives $\det(\pd^2)^{-1/2}$. Then, the corresponding action for $W$ can be expressed as
\be
S_W=-\frac{1}{2e'^2}\int d\tau d^2x \sqrt{h} ~(\gamma^2_1~ h^{\tau\tau} \pd_\tau W\pd_\tau W +\gamma^2_2~ h^{\theta\theta} \pd_\theta W\pd_\theta W+\gamma^2_3~ h^{\varphi\varphi} \pd_\varphi W\pd_\varphi W )\,.
\ee
One can always rewrite $\det(\pd^2)^{-1/2}$ as a path integral over field $\phi$, and the overall partition function can be expressed as
\bea\label{ZpW}
Z_{\phi,W} &=& \int \mD \phi~ \mD W~\exp\left[-S_{\phi,W}\right]\,.
\eea
with the effective action
\bea
S_{\phi,W} &=& \frac{1}{2e'^2}\int d\tau d^2x \sqrt{h} \nn\\
&\times&\left( h^{ab}\pd_a\phi\pd_b\phi
+\gamma^2_1~ h^{\tau\tau} (\pd_\tau W)^2 +\gamma^2_2~ h^{\theta\theta} (\pd_\theta W)^2+\gamma^2_3~ h^{\varphi\varphi} (\pd_\varphi W)^2 \right)\,.\label{Effaction}
\eea
A different way of saying rewriting $\det(\pd^2)^{-1/2}$ as a path integral over fields $\phi$ is that the path integrals with effective actions \eqref{effec3} or \eqref{Effaction} are the same. The effective action (\ref{Effaction}) can be explicitly expressed as
\be\label{SpW}
S_{\phi,W}=S_\phi+S_W\,
\ee
with
 \bea
&& S_\phi = \frac{L}{2e^2|y|^2}\int d\tau d^2x ~R^2\sin\theta~ 
\Big(\frac{F}{L R^2}\pd_\tau\phi\pd_\tau\phi + \frac{1}{R^2}\pd_\theta\phi\pd_\theta\phi+ \frac{1}{R^2\sin^2\theta}\pd_\varphi\phi\pd_\varphi\phi \Big)\,,
\nn\\
&& S_W = \frac{2}{e^2(L+r_s)}\int d\tau d^2x ~r_s^2\sin\theta ~
\Big(\frac{r_s}{\varepsilon} \frac{Lr_s^2+r_s^3}{F}\pd_\tau W\pd_\tau W +\frac{1}{r_s^2}~ \pd_\theta W\pd_\theta W+\frac{1}{r_s^2\sin^2\theta}\pd_\varphi W\pd_\varphi W \Big) \,.\nn
\eea
The two symbols $F$ and $|y|$ in the above actions are
\bea
F &=& 3r_s^2L+\frac{3}{2}r_s L^2+\frac{L^3}{3}+r_s^3\ln{L}/{\varepsilon}\,,\label{F}\\
|y| &=& \sqrt{L(L+r_s)}+r_s\text{~arcsinh} \sqrt{\frac{L}{r_s}}\,.
\eea
Equations \eqref{ZpW} and \eqref{SpW} are the partition function and the effective action that we will mainly focus on.
With that at hand, we can evaluate the partition functions of fields $\hat A_{\mu}$, $\phi$, and $W$ by the path integral.

Note that it is important to keep the original Euclidean time coordinate $\tau$ as the time coordinate for the three-dimensional theory because the periodicity of coordinate time $\tau$ is the physical inverse temperature for the observer who uses the Schwarzschild metric. If one uses different time coordinates, the physics would be difficult to discuss. For example, the low-temperature limit for the coordinate observer can be high temperatures for an observer using a different coordinate system. So we would always keep $\tau$ as our time coordinate.

\section{Behavior of different modes}
\label{behavior}

In this section, we are going to evaluate the partition function of the theory in different temperature limits. The different temperature limits are illustrated in Fig. \ref{highlowT}. The partition function for $W$ and $\phi$ will be treated separately, and we will see different behavior in different temperature limits.
The overall partition function can be written as three parts: bulk contribution $Z_{\hat A}$ multiplied by $Z_\phi$ and $Z_W$
\bea
Z &:=&  Z_{\hat A}\times Z_{\phi}\times Z_{W}\\
&=& \int \mD \hat A_\mu ~e^{-\frac{1}{4e^2}\int_\mM d\tau d^3x\sqrt{g}~ \hat F^{\mu\nu}\hat F_{\mu\nu}}\times \int \mD \phi ~e^{-S_{\phi}}\times \int \mD W~ e^{-S_{W}}\,.\nn
\eea
We are going to calculate the partition function for the bulk fluctuation modes $Z_{\hat A}$ in the next subsection. After that, we will evaluate $Z_\phi$ and $Z_W$ in the high- and low-temperature limits in Secs. \ref{BHphi} and \ref{BHN}.

\subsection{Bulk fluctuation modes}
\label{Bulkfm}

 Now, let us evaluate the entropy of bulk fluctuation modes. The details of the calculation are exiled to Appendix \ref{Sch}, to avoid being distracted from the main text. This is the standard blackbody calculation on curved spacetime. For electromagnetism, we have two physical polarizations in the bulk, which are both bosonic and massless.
The bulk partition function reads as
\be\label{lnZF}
\ln Z_{\hat A} = -\frac{4\pi^3}{45}\frac{1}{\beta^3}\frac{r_s^4}{\varepsilon} -\frac{16\pi^3}{45}\frac{r_s^3}{\beta^3}\ln \frac{L}{\varepsilon}  -\frac{4\pi^3}{45}\frac{1}{\beta^3}\left( -\frac{r_s^4}{L}+6~r_s^2 L+2~r_s L^2+\frac{L^3}{3}\right)\,.
\ee
and the corresponding entropy can be written as
\bea
\mS_{\hat A} &=&  \frac{16\pi^3}{45}\frac{1}{\beta^3}\frac{r_s^4}{\varepsilon}+\frac{64\pi^3}{45}\frac{r_s^3}{\beta^3}\ln \frac{L}{\varepsilon}
 +\frac{16\pi^3}{45}\frac{1}{\beta^3}\left( -\frac{r_s^4}{L}+6~r_s^2 L+2~r_s L^2+\frac{L^3}{3}\right)\,.\label{bulkSS}
\eea

Let us look at the first two terms
\be\label{firstS}
\mS_{0}=\frac{16\pi^3}{45}\frac{1}{\beta^3}\frac{r_s^4}{\varepsilon}+\frac{64\pi^3}{45}\frac{r_s^3}{\beta^3}\ln \frac{L}{\varepsilon}\,.
\ee
First of all, those contributions cannot be seen in the extremal black hole (or superlow temperature) case where we have $\beta\to \infty$ while keeping the radius of the black hole to be finite.
For a finite temperature black hole, defining the proper distance from the real horizon to the stretched horizon as $\delta$, we have
\bea\label{delta}
\delta &=& \int_{r_s}^{r_s+\varepsilon} \sqrt{g_{rr}}~dr
\approx  2\sqrt{\varepsilon~r_s }\,.
\eea
Thus, we have $\delta^2 \approx 4\varepsilon ~r_s$. For the finite temperature black hole, with the inverse temperature
\be
\beta =8\pi G_N M\,,
\ee
the entropy (\ref{firstS}) can be written as
\be
\mS_{0}\propto \frac{r_s^2}{\delta^2}+\ln \frac{L r_s }{\delta^2}\,.\label{ententropy}
\ee
Those contributions arise because of the redshift between the horizon and the coordinate observer sitting at infinity. Any finite frequency modes near the horizon would have zero frequency as seen by a coordinate observer. We have an infinite number of states with zero energy, and summing over all of those states with a UV cutoff would give us the above result. Those very dense ground states mainly come from the near-horizon region; thus we have an area contribution and corrections \cite{Susskind2004}.
This result was used to understand the Bekenstein-Hawking entropy by some authors, for example \cite{tHooft:1984kcu}, and it was also interpreted as the entanglement entropy across the stretched horizon \cite{Srednicki:1993im}. \footnote{However there are different opinions. For instance, Susskind showed that the contribution should be absorbed by the renormalization of Newton's constant $G_N$ due to the loop contribution \cite{Susskind:1994sm}.}

The other terms
\bea
\mS_{\hat A} = \frac{16\pi^3}{45}\frac{1}{\beta^3}\left( -\frac{r_s^4}{L}+6~r_s^2 L+2~r_s L^2+\frac{L^3}{3}\right) \propto \text{Volume}\times T^3\,\label{ShatA}
\eea
are entropy of the thermal fluctuation modes of gauge fields living in the bulk, which can be regarded as the curved spacetime analog of the blackbody result.
$S_{\hat A}$ is finite and more or less proportional to the volume between the two boundaries multiplied by $T^3$.

\subsection{Zero longitudinal momentum modes of $A_r$}
\label{BHphi}

In this section, we evaluate the partition function for $\phi$, which is zero longitudinal momentum modes of $A_r$. The partition function $Z_{\phi}$ can always be written as
\be
Z_\phi = \int \mD \phi ~e^{-S_\phi}\,,\label{Zphi}
\ee
with the action
\be
S_\phi = \frac{L}{2e^2|y|^2}\int d\tau d^2x ~R^2\sin\theta~ \Big(\frac{F}{L R^2}\pd_\tau\phi\pd_\tau\phi + \frac{1}{R^2}\pd_\theta\phi\pd_\theta\phi+ \frac{1}{R^2\sin^2\theta}\pd_\varphi\phi\pd_\varphi\phi \Big)\,,\label{SSphi}
\ee
with
\bea
F = 3r_s^2L+\frac{3}{2}r_s L^2+\frac{L^3}{3}+r_s^3\ln{L}/{\varepsilon}\,,~~~|y| = \sqrt{L(L+r_s)}+r_s\text{~arcsinh} \sqrt{\frac{L}{r_s}}\,.\label{Fy}
\eea
The action (\ref{SSphi}) has different behavior in different temperature limits, and we will see how the entropy contribution from $\phi$ changes as we vary the temperature of the system.
One obvious thing to notice is that there is a transition of dominance in function $F$ shown in Eq. \eqref{Fy} at different temperatures. 
The $L^3$ term is the dominant contribution for high temperatures $L\gg r_s$, and the UV cutoff $\varepsilon$ is absent in $S_\phi$. 
While at low temperatures $L\ll r_s$, the most important term is $r_s^3\ln L/\varepsilon$, and we will see different behavior of the corresponding entropy. 
We always assume that the short distance cutoff $\varepsilon$ is always much smaller than $L$ and $r_s$.

\subsubsection{High-temperature limit}
\label{high}

Let us first discuss the high-temperature limit $r_s\ll L$, as shown in the first panel of Fig. \ref{highlowT}. If
\be
r_s^3\ln\frac{L}{\varepsilon}\ll L^3\,
\ee
is satisfied, then $L^3$ is the most important term in $F$. In this case, $F$ can be approximated as
\be
F \approx \frac{L^3}{3}\,.
\ee
Function arcsinh$(x)$ is always much smaller than $x$ for large value of $x$, so $|y|$ can be written as
$|y|\approx L$.
Then, the effective action in the high-temperature limit can be expressed as
\bea
S_{\phi} &=& \frac{1}{2e^2L}\int d\tau d^2x ~L^2\sin\theta~ \Big(\frac{1}{3}\pd_\tau\phi\pd_\tau\phi + \frac{1}{L^2}\pd_\theta\phi\pd_\theta\phi+ \frac{1}{L^2\sin^2\theta}\pd_\varphi\phi\pd_\varphi\phi \Big)\,. \label{highS}
\eea
The effective action for $\phi$ is just a scalar field living on a $ {S}^1\times {S}^2$, where the length scale of $ {S}^1$ is $\beta$, and the length scale of $ {S}^2$ is $L$. 

Equipped with the effective action (\ref{highS}), we can directly calculate the partition function $Z_{\phi}$ shown in (\ref{Zphi}).
We can absorb the finite constant $1/3$ in front of $\pd_\tau\phi\pd_\tau\phi$ term in the action into the redefinition of $\tau$ to $\tau'$.
Let us suppose the fluctuation modes of $\phi$ are
\bea
\phi(x^a)&=&  \mathcal N_\phi \cdot \sum_\omega \sum_{l,m}e^{i\omega\tau'}Y_{lm}(\theta,\varphi)\tilde{\phi}(\omega,l,m)\,,
\eea
where $\mathcal N_\phi$ is a normalization constant.
The partition function for $\phi$ in the canonical ensemble can be written as
\be
\ln Z_\phi = -\sum_\omega \ln(1-e^{-\beta' \omega})\,.
\ee
The calculation is similar to the bulk fluctuation modes calculation demonstrated in Appendix \ref{Sch}. We can change the summation of $\omega$ to integration by introducing density of state $g(\omega)$, which gives out
\bea\label{ZFphi}
\ln Z_\phi &=& -\int_0^\infty g(\omega)\ln(1-e^{-\beta' \omega})d\omega
=
 \beta'\int_0^\infty\frac{\Gamma(\omega)}{e^{\beta' \omega}-1}d\omega\,.
\eea
$\Gamma(\omega)$ is defined by $d\Gamma =g(\omega)d\omega$.
According to the dispersion relation in this background, $\Gamma(\omega)$ can be counted as
\bea
\Gamma &=& \sum_{l,m}\frac{\beta'}{2\pi}\sqrt{\frac{l(l+1)}{L^2}}=\frac{\beta'}{2\pi}\sum_l(2l+1)\sqrt{\frac{l(l+1)}{L^2}}\,.
\eea
The summation is from $l=0$ to the level with energy $\omega$. Putting everything back into Eq. (\ref{ZFphi}) and changing the summation of $l$ into the integral, we can write the partition function as
\bea
\ln Z_\phi
&=& \frac{\beta'^2}{2\pi}L^2\int_0^\infty \frac{d\omega}{e^{\beta' \omega}-1} \int_l d\left(\frac{l(l+1)}{L^2}\right)~\sqrt{\frac{l(l+1)}{L^2}}\,.
\eea
$\sqrt{l(l+1)/L^2}$ is exactly the energy carried by a particle with angular momentum $l(l+1)$. Thus we can change a variable and get
\be
\ln Z_\phi
= \frac{\beta'^2 L^2}{2\pi}\int_0^\infty \frac{d\omega}{e^{\beta' \omega}-1} \int_0^{\omega^2} dx~\sqrt{x}=\frac{\beta'^2 L^2}{3\pi}\int_0^\infty \frac{\omega^3 d\omega}{e^{\beta' \omega}-1}\,.
\ee
Then the above expression can be worked out, and the logarithm of the partition function can be expressed as
\be
\ln Z_\phi =\frac{\pi^3}{45}\frac{L^2}{\beta'^2}=\frac{\pi^3}{135}\frac{L^2}{\beta^2}\,.
\ee
Then, the corresponding entropy can be written as
\be
\mS_{\phi}=\frac{\pi^3}{45}\frac{L^2}{\beta^2}\,.\label{phihigh}
\ee
The above result can be compared with the entropy of $\phi$ in the flat case in \cite{Cheng2022}.
Thus, we can conclude that the zero modes $\phi$ share similar properties with the flat case in the high-temperature limit $r_s\ll L$.

So the high-temperature behavior of field $\phi$ is just like a three-dimensional scalar field with inverse temperature $\beta$, whose entropy is standard, as shown in Eq. (\ref{phihigh}). There is not much difference with the flat case.

\subsubsection{Low-temperature limit}
\label{low}

As the temperature becomes superlow, $r_s$ can be larger than $L$, and we arrive at the low-temperature limit of the system with $r_s\gg L$, as shown in the second panel of Fig. \ref{highlowT}. 
As we can see, there is a transition of dominance between different terms in $F$, and in the low-temperature limit $F$ can be approximated as
\be
F\approx r_s^3\ln\frac{L}{\varepsilon}\,.
\ee
We also have $|y|^2\approx L(L+r_s)$. Then, the effective action for $\phi$ in the low-temperature limit can be written as
\bea
S_{\phi} &=& \frac{1}{2e^2(L+r_s)}\int d\tau d^2x ~r_s^2\sin\theta \Big(\frac{r_s}{L}\ln \frac{L}{\varepsilon}\pd_\tau\phi\pd_\tau\phi + \frac{1}{r_s^2}\pd_\theta\phi\pd_\theta\phi+ \frac{1}{r_s^2\sin^2\theta}\pd_\varphi\phi\pd_\varphi\phi \Big)\,.\label{SSphiphi}
\eea
As discussed before, the time of the three-dimensional action should always be chosen as the coordinate time $\tau$.
Doing so, the coefficient in front of $(\pd_\tau\phi)^2$ is fixed and always much larger than 1. Especially when we take the real event horizon limit, i.e. the $\varepsilon \to 0$ limit, the path integral localizes on the space of zero energy modes $\pd_\tau \phi=0$.
Then the partition function of the field $\phi$ only depends on the radius of ${S}^2$.

Because of the localization discussed above, the logarithm of the partition function for the zero-point energy is no longer linear in $\beta$ but only depends on the radius of $S^2$ and UV cutoff, which can be expressed as
\be
\ln Z_\phi\propto r_s^2 \cdot \Lambda^2\,.
\ee
Let us assume the UV cutoff is the Planck scale, then the corresponding entropy is of Bekenstein-Hawking entropy magnitude
\be
\mS\propto\frac{r_s^2}{l_p^2}\,.\label{philow}
\ee
So, we can say that, in the low-temperature limit, the entropy of $\phi$ can be written as the area of the horizon divided by the Planck area $l_p^2$.

\begin{figure}
  \centering
  \includegraphics[width=0.6\textwidth]{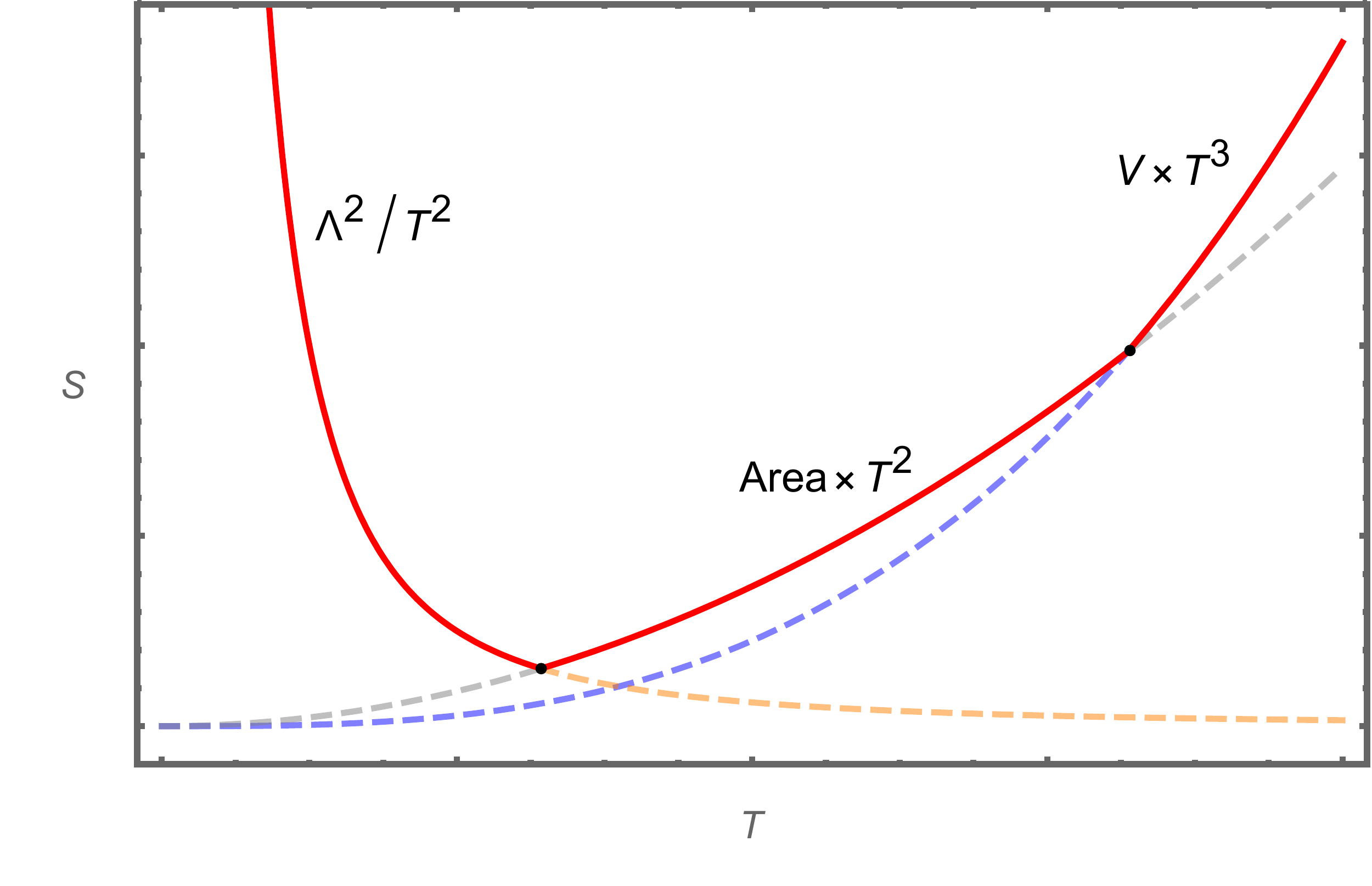}\\
  \caption{Sketch of the entropy of the gauge theory as we vary the temperature of the black hole. The high-temperature behavior is qualitatively like the flat case as shown in Fig. \ref{boxphase}. At superlow temperatures, the entropy is proportional to the area of the horizon, multiplied by the UV cutoff $\Lambda$ squared. }\label{BHphase}
\end{figure}

Now, let us summarize what we have got for the entropy of zero longitudinal momentum modes $\phi$. The entropy of the whole system is illustrated in Fig. \ref{BHphase}. 
At high temperatures, $L\gg r_s$, the entropy of $\phi$ is more or less like what we have in the flat case, namely the area of the box times temperature squared $L^2\times T^2$. Hence, the presence of a small black hole merely gives a temperature to the system. This amount of entropy then has competition with the bulk fluctuation modes. 
At very high temperatures, the volume times temperature cubed wins, and the area times temperature squared wins at relatively lower temperatures. 
This is more or less the same story told in \cite{Cheng2022}, reviewed in Sec. \ref{review}. 
The difference starts to show up at superlow temperatures; one can compare Figs. \ref{BHphase} and \ref{boxphase} to see the difference. In the black hole case, the entropy for $\phi$ is shown in Eq. (\ref{philow}) at superlow temperature, which is reminiscent of the Bekenstein-Hawking entropy. In comparison, this phenomenon cannot be seen in the flat case, where the entropy has a logarithm behavior as shown in Fig. \ref{boxphase}.

\subsection{Boundary-stretched Wilson lines}
\label{BHN}

Now, let us also discuss the different behavior of field $W$ in different temperature limits in this subsection. The effective action for $W$ is
\bea
S_W &=& \frac{2}{e^2(L+r_s)}\int d\tau d^2x ~r_s^2\sin\theta 
\Big(\frac{r_s}{\varepsilon} \frac{Lr_s^2+r_s^3}{F}\pd_\tau W\pd_\tau W +\frac{1}{r_s^2}~ \pd_\theta W\pd_\theta W+\frac{1}{r_s^2\sin^2\theta}\pd_\varphi W\pd_\varphi W \Big) \,.\nn\\~\label{SSNN}
\eea
with
\be
F= 3r_s^2L+\frac{3}{2}r_s L^2+\frac{L^3}{3}+r_s^3\ln{L}/{\varepsilon}\,.
\ee
The infinite coefficient introduces a localization on the zero energy modes $\pd_\tau W=0$, and the entropy of the Boundary-stretched Wilson lines $W$ is more or less
\be
\mS\propto\frac{r_s^2}{l_p^2}\,,\label{entropyN}
\ee
for the same reason discussed in the previous subsection. 

Only when the size of the box $L$ is extremely large compared to $r_s$, namely the high-temperature limit, some power of $r_s/L$ may overcome $r_s/\varepsilon$, and we get a finite coefficient.
Then, the corresponding entropy can be expressed as
\be
\mS\propto r_s^2 \times T^2\,,
\ee
which is a finite constant.
In this limit, the entropy from $W$ can always be ignored, because the entropy of fluctuation modes always plays the dominant role at high temperatures.

There can also be constant modes, and topological modes due to the map between the Euclidean time circle and the U(1) group. The entropy of the constant modes and winding modes of $W$ are never comparable with (\ref{entropyN}), but contribute as logarithm corrections. We have
\be
Z_0=\sqrt{\frac{e^2\beta(L+r_s)}{\A}}\,,~~~~~~~~~~~
Z_w=\sum_n e^0=\zeta(0)\,,\label{Z0w}
\ee
which contribute as logarithms of the temperature $T$, the coupling constant $e$, and other length scales. 
Note that the constant modes always contribute a logarithm term in entropy, but the winding modes tend to be less important as can be seen from \eqref{Z0w}. This is because for the superlow temperatures, $\beta$ is superlarge and we can only see the mode of the zero winding number. The winding modes contribution in relatively high temperatures is more or less the same as the flat case \cite{Cheng2022}, which is not expatiated here because those modes are not comparable with the fluctuation modes at high temperatures.
The overall logarithm correlation from constant and winding modes is an important aspect of the entropy but is small compared which other contributions. 
So, we can say that the entropy of $W$ follows the same qualitative pattern as the entropy behavior shown in Fig. \ref{BHphase}.


Let us briefly summarize this section. Basically, we have three different phases. At high temperatures, the bulk fluctuation modes shown in Sec. \ref{Bulkfm} are the most important contribution. The entropy is proportional to the volume of the box multiplied by the temperature cubed, as shown by the right part of the curve in Fig. \ref{BHphase}. We also have a contribution (\ref{ententropy}) scale as the area of the horizon plus logarithm correction, contained in the bulk fluctuation contribution. As the temperature cools down, the boundary area contribution dominates over other contributions. The corresponding entropy is shown in Eq. (\ref{phihigh}). At superlow temperatures, we have an entropy of Bekenstein-Hawking magnitude
\be\label{secondS}
\mS\propto\frac{r_s^2}{l_p^2}\,,
\ee
coming from both the zero modes $\phi$ and the Wilson lines $W$. This contribution will not die off at zero temperature. So we expect the behavior would persist for the extremal black holes. It is interesting to further verify this point for the extremal black hole case.

\section{Extremal black hole}
\label{extremal}

The entropy in the superlow temperature limit of finite temperature black holes is because of the localization of modes on a specific region in the phase space. 
Although we inferred that the mechanism should be responsible for the entropy of extremal black holes, a calculation on an extremal black hole background is vital for the story. The main task of this section is to show the same localization mechanism also works for the extremal black hole case.
We are going to evaluate the partition function and entropy of U(1) gauge theory with nontrivial boundary conditions on multiple boundaries on a (near-)extremal black hole background. 

To get an extremal black hole, we start with the Euclidean Reissner-Nordstr\"{o}m metric, which can be written as
\be
ds^2=f(r)d\tau^2+f(r)^{-1}dr^2+r^2d\Omega^2_2\,,
\ee
with
\be
f(r)=1-\frac{2G_N M}{r}+\frac{G_N Q^2}{r^2}\,.
\ee
The horizon is a null surface, which can be obtained by solving $g_{rr}=0$. The inner and outer horizons can be written as
\be
r_{\pm}=(G_N M\pm \sqrt{(G_N M)^2-G_N Q^2})\,.
\ee
The extremal limit is the limit where we have a double zero, which means
\be
r_+=r_-=G_N M\,.
\ee
Thus, the metric of an extremal black hole can be written as
\be\label{EBH}
ds^2=\left(1-\frac{r_H}{r}\right)^2 dt^2+\left(1-\frac{r_H}{r}\right)^{-2}dr^2+r^2d\Omega_2^2\,,
\ee
with
\be
r_H=G_N M\,.
\ee
We can give this black hole system a very tiny temperature by replacing the double zero at the horizon with two single zeros $r_{\pm}=r_H\pm \varepsilon$ with the short distance cutoff $\varepsilon$. 
The inverse temperature of the near extremal black hole can be expressed as 
\be\label{beta}
\beta=\frac{2\pi r_H^2}{\varepsilon}\,,
\ee
which is infinity in the limit $\varepsilon \to 0$.

Now, let us put U(1) gauge theory with boundaries at the stretched horizon and at location $r_H+L$ in spacetime \eqref{EBH}. The original Euclidean action can be written as
\be\label{action2}
S_E=\frac{1}{4e^2}\int_{\mathcal{M}}d^4x \sqrt{g} ~F^{\mu\nu}F_{\mu\nu}\,.
\ee
We are going to focus on the first boundary condition \eqref{first}, where we allow residual $A_r$ modes on boundaries
  \be
  \delta A_{a}\Big{|}_{\pd \mM}=0,~~~~~~\delta A_r\Big{|}_{\pd \mM}=f(x^a)\,.
  \ee
$f(x^a)$ can have local dependence of $x^a=(\tau,\theta,\phi)$ on different boundaries. $A_{a}\big{|}_{\pd \mM}$ are fixed configurations, which can be set to zero. 

Following the same logic as the finite temperature case, we shift the radius direction by $r_H$
\be
\rho=r-r_H\,,
\ee
and then define the proper distance $y$ along the radius direction as
\be
dy=\frac{\rho+r_H}{\rho}d\rho\,.
\ee
Solving the above equation, the proper distance $y$ can be expressed as
\be
y=\rho+ r_H\ln \frac{\rho}{r_H}\,.
\ee
$y$ takes value from $y_1$ to $y_2$, with
\be
y_1 = \varepsilon+{r_H}\ln \frac{\varepsilon}{r_H}\,,~~~~~y_2 = L+{r_H}\ln \frac{L}{r_H}\,.\\
\ee
We have ln$(0)=-\infty$ and thus $y_1\to-\infty$\footnote{This is different with the finite temperature case. In the finite temperature case, the left boundary is localized at $y_1\to0$ because of arcsinh$(0)=0$.}. So $|y|$ can be approximated as
\be
|y|=y_2-y_1\approx L+r_H\ln\frac{L}{\varepsilon}\,.
\ee

On the stretched horizon, the residual degrees of freedom of $A_y$ are set to $\phi(x^a)/L$, and the field $A_y$ can be decomposed as
\be\label{decom}
A_y(x^{\mu})=\hat{A}_y(x^\mu)+\frac{\phi(x^a)}{|y|}\,;
\ee
the difference between the two boundaries is captured by the Wilson lines stretched between the two boundaries at $y=y_1$ and $y=y_2$,
\be
W(x^a)=i\int_{y_1}^{y_2}dy ~\hat A_y.
\ee
Thus, we have divided our gauge fields into three different parts: the bulk fluctuation modes $\hat A_y$ that vanish on both boundaries, zero longitudinal momentum modes $\phi(x^a)$, and boundary-stretched Wilson lines $W(x^a)$. The symplectic form and commutation relations of those modes were worked out by carefully analyzing the phase space of the theory in \cite{Cheng2022}. So one can just put those modes in the Euclidean path integral to calculate the corresponding partition function and entropy of them.

The partition function can be formally written as
\be\label{Z}
Z=\int \mD \hat A_\mu ~\mD \phi ~\mD W~\exp\left\{-S_E[\hat A_\mu, \phi, W]\right\}\,.
\ee
The action $S_E$ can be separated into 
\be
S_E=S_{\hat A}+S[\phi,W]\,,
\ee
where
\be\label{SA}
S_{\hat A}=\frac{1}{4e^2}\int_\mM d\tau d^3x\sqrt{g}~ \hat F^{\mu\nu}\hat F_{\mu\nu}\,.
\ee
The bulk fluctuation modes reviewed in the previous section will not survive, because of the finite $r_H$ and the tiny temperature $T=1/\beta\propto \varepsilon$. So we will mainly focus on fields $\phi$ and $W$ to see the behavior of those modes. Actually, all of the fluctuation mode contributions whose entropy is proportional to the temperature will not play any role in the final result, and we expect to see entropy contributions that look like the Bekenstein-Hawking entropy for extremal black holes.

Ignoring the bulk fluctuation modes, we are going to focus on fields $\phi$ and $W$, whose action can be written as
\bea
S[\phi,W] = \frac{1}{2e^2 |y|^2}\int_\mM d\tau d^3x \sqrt{g} \left[ g^{ab}\pd_a\phi\pd_b\phi\right]
-\frac{i}{e^2|y|}\int d\tau d^2x\left[\sqrt{g} ~ g^{ab}\right]_{y=y_1}^{y=y_2}\pd_a W \pd_b \phi\,.
\eea
Note that the action $S[\phi,W]$ contains four-dimensional and three-dimensional integrations, while $\phi(x^a)$ and $W(x^a)$ only depend on transverse directions along the boundaries. So we can dimension reduce the action $S[\phi,W]$ to three-dimensional and work out an effective action for $\phi$ and $W$. 
Following a similar dimensional reduction procedure in the previous section, we obtain the three-dimensional effective action for fields $\phi$ and $W$
\be
S_{\phi,W}=S_\phi+S_W\,,\label{phiWaction}
\ee
with
\bea
S_\phi &=& \frac{L}{2e^2|y|^2}\int d\tau d^2x~r_H^2\sin \theta
\left( \frac{r_H^2}{\varepsilon L}\pd_\tau\phi\pd_\tau\phi+\frac{1}{r_H^2}\pd_\theta\phi\pd_\theta\phi+ \frac{1}{r_H^2\sin^2\theta}\pd_\varphi\phi\pd_\varphi\phi \right)\,,\label{Sphi}\\
S_W &=& \frac{2L}{e^2(r_H+L)^2}\int d\tau d^2x~r_H^2\sin \theta\nn\\
&~&~~~~~~~~~~~~~~~\times\left( \frac{(r_H+L)^2}{\varepsilon L}\pd_\tau W\pd_\tau W+\frac{1}{r_H^2}\pd_\theta W\pd_\theta W+ \frac{1}{r_H^2\sin^2\theta}\pd_\varphi W\pd_\varphi W \right)\,.\label{SW}
\eea
Note that one of the important ingredients in the above procedure is to keep the original coordinate time $\tau$ as the time coordinate for the three-dimensional effective actions. This is because we always need to stick to the observer with the inverse temperature \eqref{beta}, otherwise, the physics can be completely different.

As we can see from the effective actions \eqref{Sphi} and \eqref{SW}, for finite $r_s$ and $L$, the coefficients in front of terms $(\pd_\tau\phi)^2$ and $(\pd_\tau W)^2$ are proportional to $1/\varepsilon$. Taking the $\varepsilon \to 0$ limit, the coefficients are divergent, which introduces a localization on the space of $\pd_\tau\phi=\pd_\tau W=0$ in the path integral \eqref{Z}.
The localization is similar to what we discussed in the low-temperature limit of the finite-temperature black hole case, as reviewed in the previous section. Then the entropy of those modes can be calculated as
\be
S\propto\frac{r_H^2}{l_p^2}\,,\label{extremalS}
\ee
which is the same result as the superlow temperature entropy \eqref{secondS} in the finite-temperature black hole case. 
The partition functions of the constant modes and winding modes of $W$ can be calculated, which is the logarithm of the length scales in the theory. One thing worth noticing is that the entropy of constant modes proportional to
\be
S_0\propto\ln|y|\,,
\ee
can be a large contribution.

Note that in the (near-)extremal black hole case, the localization of $\pd_\tau\phi=\pd_\tau W=0$ space in the path integral is very straightforward as shown in the actions \eqref{Sphi} and \eqref{SW}. This is different from the finite-temperature black hole case where we have to take different temperature limits and the localization only shows up in the superlow temperature limit. 
So the localization mechanism and the entropy (\ref{extremalS}) and logarithm corrections are intrinsic for the (near-)extremal black hole and only capture the low-temperature properties of the black hole system.
So the localization in $\pd_\tau\phi=\pd_\tau W=0$ space should be used as a general mechanism to explain the entropy of the extremal black hole in a box. Moreover, one of the surprising properties of entropy is that the modes are effectively living near the horizon rather than the other boundary. So the entropy \eqref{extremalS} only depends on $r_H$. There might be redshift-related arguments, but we do not have a good explanation here.

As for where the entropy \eqref{extremalS} comes from, the zero-point energy of boundary modes contributes as 
\be
\beta \text{Area} \times \Lambda^3 \in \ln Z
\ee
in the logarithm of the partition function. This part has no contribution to the entropy because it is linear in $\beta$. $\pd_\tau\phi=\pd_\tau W=0$ means there is no $\beta$ dependence in this part and correspondingly this part contributes to the entropy as $\text{Area} \times \Lambda^2$. There must be fundamental explanations of those modes from symmetry-related arguments. And, the microscopic derivations for the Bekenstein-Hawking entropy of extremal black holes \cite{Strominger1998, Birmingham1998, Guica2009, Carlip2011,Brown1986, Solodukhin1999, Carlip1999a, Carlip1999, Silva2002, Carlip2007, Castro2010, Wang2010, Chen2010, Carlip2011a, Compere2017, Song2012, Carlip2018, Carlip2019, Strominger1996, Callan1996, Peet2000, Das2001, Maldacena1996, Emparan2006} might give the right physical explanation of those modes.

So, we have confirmed that the localization on $\pd_\tau\phi=\pd_\tau W=0$ space in the path integral discovered in the superlow temperature limit discussed in the previous section persists in the (near-)extremal black hole case. The mechanism is the main reason we get Bekenstein-Hawking-like entropies from (near-)extremal black holes.

\section{Conclusion and discussion}
\label{conclusions}

This paper evaluates the partition function and entropy of U(1) gauge theory on black hole background with nontrivial boundary conditions, using the Euclidean path integral method. 
The allowed physical degrees of freedom are the bulk fluctuation modes, the zero longitudinal momentum modes of $A_r$, and the boundary-stretched Wilson lines. The effective actions for different modes are derived, and we can separately calculate the entropy contributions from different modes.

The high-temperature limit of the black hole is fairly similar to the flat case reviewed in Sec. \ref{review}, where the presence of the black hole merely gives a temperature to the system.
The dominant contribution comes from the bulk fluctuation modes, whose entropy is shown in Eq. \eqref{bulkSS}. 
Most of the entropy of the fluctuations is proportional to the volume of the region where the semiclassical fields live, multiplied by the temperature cubed.
There is also a contribution from the modes living very close to the horizon, whose entropy is proportional to the area of the horizon plus logarithm corrections.
Interesting phenomena start to happen at relatively lower temperatures. 
As we gradually decrease the temperature, bulk fluctuation modes became less important, and the entropy of the zero modes $\phi$ and Wilson lines $W$ behave as the area of the box multiplied by the temperature squared are the most important contribution as shown in Fig.  \ref{BHphase}.
For superlow temperature, the entropy of $\phi$ and $W$ scale as the area of the horizon divided by UV cutoff squared. This is because the coefficients in front of $(\pd_\tau \phi)^2$ and $(\pd_\tau W)^2$ in the effective actions diverge,
introducing localizations in the space of zero-energy modes $\pd_\tau \phi=\pd_\tau W=0$ in the Euclidean path integral.
Then entropy of those modes naturally is proportional to the horizon area divided by the Planck area. We also have extra logarithm corrections from the constant modes and topological modes of $W$.
The localization mechanism of the U(1) gauge theory is also confirmed for the extremal black holes. 
The overall behavior of the entropy is depicted in Fig. \ref{BHphase}, and we can see the transitions of dominances between the low-temperature black hole and high-temperature ones clearly.

Now, the question is how to understand the large entropy \eqref{philow} we have gotten for the low-temperature black hole. Let us take $\phi$ as an example to show where this large entropy comes from. The finite temperature partition function for a three-dimensional massless bosonic field can be written as
\be
\ln Z_\phi \propto \beta \A \cdot \Lambda^3+ \A \int d^2p~\ln(1-e^{-\beta p})\,,
\ee
where $\Lambda$ is the UV cutoff and $p=\sqrt{\vec p^{~2}}$. The first part proportional to the volume of the whole spacetime divided by the smallest volume unit is the zero-point energy of the field theory, which is a constant piece in the free energy. Constant free energy does not contribute to entropy, and thus can be ignored.
The second part is finite and gives out the $\A \times T^2$ contribution in entropy. However, the finite contribution cannot be seen at superlow temperatures.
Now because of the localization, the zero-point energy of the field is not a constant free energy anymore. The logarithm of the partition function only depends on the area of the boundary because $\phi$ only depends on the spatial coordinates on the boundary. So the partition function of the zero-point energy can be written as
\be
\ln Z_\phi\propto \A \cdot \Lambda^2=\frac{\A}{l_p^2}\,,\label{introlnZ}
\ee
if we suppose the UV cutoff is at the Planck scale. The corresponding entropy is of Bekenstein-Hawking entropy magnitude. There are also constant modes and winding mode contributions at superlow temperatures, whose entropy scales as the logarithm of the coupling constant and other length scales. It is interesting to notice that the area shown in Eq. \eqref{introlnZ} only contains the area of the horizon not the area of the other boundary.

The above argument suggests that the low-temperature Bekenstein-Hawking entropy might come from the zero modes $\phi$ and boundary-stretched Wilson lines $W$. 
For the transitions of dominances for the finite temperature black hole, especially at the low temperature, there should be some symmetry-breaking pattern to explain them. 
As shown in Fig. \ref{boxphase}, the low-temperature entropy of the flat case scales as a logarithm because the degeneracy manifold is a circle in the symmetry-breaking phase of U(1) global symmetry. 
This can be seen from the bottom of the ``Mexican hat" potential. 
In such a sense, the low-temperature Bekenstein-Hawking-like entropy
\be
\mS\propto\frac{\A}{l_p^2}\nn\,,
\ee
can also be explained from a symmetry-breaking viewpoint. 
We are sitting in a global symmetry-breaking phase at low temperature with degeneracy $\exp\mS$. 
Note that those amounts of entropy come from the modes in the limit
\be
\lim_{\omega\to 0} ~\tilde{\phi}(\omega, x^2,x^3)e^{i\omega\tau}\,
\ee
because of the localization on $\pd_\tau \phi=0$ space. This might can be thought of as calculating the entropy of the soft hair of the black hole system \cite{Hawking:2016sgy, He:2014laa, Haco:2018ske, Hawking:2016msc, Strominger2018, Cheng:2020vzw, Cheng:2022xyr, Cheng:2022xgm}.  The exact relation between our story and the soft hair of black holes needs further studies.

Note that we have gotten two Bekenstein-Hawking-like entropies in this paper.
The original point of this paper is that the zero modes $\phi$ and the Wilson lines $W$ have an entropy proportional to the horizon area divided by the Planck area for superlow temperature black holes, which might be used to understand the Bekenstein-Hawking entropy because of the right magnitude.
Note that there is also part of the entropy (\ref{ententropy}) in the bulk fluctuation modes for the finite temperature black hole system, which is also proportional to the area divided by the UV cutoff squared. This entropy was used to understand the microscopic degrees of freedom of finite-temperature black hole systems \cite{tHooft:1984kcu}, and was interpreted as the entanglement entropy across the stretched horizon \cite{Srednicki:1993im}.
If we accept the above argument, we might tend to interpret that the finite-temperature Bekenstein-Hawking entropy comes from some extra microscopic structure near the horizon, like entanglement across the horizon, but the (near-)extremal black hole entropy comes from different places as the finite-temperature black hole. 
The (near-)extremal black hole entropy only appears in low temperatures and comes from the breaking of global symmetries. 
Thus, we have two different types of Bekenstein-Hawking-like entropies for finite-temperature and (near-)extremal black holes, and they both behave like the area of the horizon divided by the UV cutoff squared.
We leave the symmetry-breaking explanation of the phase transitions and other related topics for further study.


\begin{acknowledgments}
We would like to thank Ankit Aggarwal, Jan de Boer, Diego Hofman, and Pujian Mao for their useful discussions. 
This work is supported by the National Natural Science Foundation of China (NSFC) under Grants No. 11905156 and No. 11935009. 
\end{acknowledgments}

\appendix

\section{Would-be gauge degrees of freedom}
\label{solB}

In this appendix, let us briefly discuss the physics related to the Neumann-like boundary condition \eqref{Neumann}, where the boundary would-be gauge degrees of freedom are allowed. \footnote{A more careful analysis of those modes can be found in appendix A of \cite{Cheng2022}}

The boundary configurations that respect \eqref{Neumann} are the flat boundary configurations.
We can find bulk modes $B_\mu$ that correspond to those boundary configurations and integral over those modes in the Euclidean path integral to evaluate the partition function.
So, the main task in this appendix is to find a solution to the bulk equation of motion on the Euclidean Schwarzschild background.
To avoid complicating the story, we can solve the problem using the original metric (\ref{metricrho})
\be
ds^2=\frac{\rho}{\rho+r_s}d\tau^2+ \frac{\rho+r_s}{\rho}d\rho^2 +(\rho+r_s)^2 d\Omega^2\,.
\ee
After getting the solutions, we can perform a coordinate transformation from $x^{\mu}=(\tau,\rho,\theta,\varphi)$ to $x^{\mu'}=(\tau,y,\theta,\varphi)$ to obtain the solutions on the infalling coordinate system.

In coordinate system (\ref{metricrho}), the problem reads as
\bea\label{solBeq}
 \nabla_\mu F^{\mu\nu}_{(B)} &=& 0\,\nn\\
B_a \big{|}_{\rho=\varepsilon}&=&f_a^{(l)}(x^a)\,\\B_a \big{|}_{\rho=L}&=&f_a^{(r)}(x^a)\nn\,.
\eea
Now, we use $f_a^{\alpha}(x^a)$ with $\alpha=(r)$ or $(l)$ to denote the boundary configuration of $B_{a}$. $B_{\rho}$ should change accordingly. We will fix the boundary configurations to be flat later. The bulk equation of motion can be further written as
\bea
\nabla_{\mu}F_{(B)}^{\mu\nu} &=& \pd_{\mu}F_{(B)}^{\mu\nu}+\Gamma^{\mu}_{\mu\lambda}F_{(B)}^{\lambda\nu}+\Gamma^{\nu}_{\mu\lambda}F_{(B)}^{\mu\lambda}
=\pd_{\mu}F_{(B)}^{\mu\nu}+\Gamma^{\mu}_{\mu\lambda}F_{(B)}^{\lambda\nu}
\newe \frac{1}{\sqrt{g}}\partial_{\mu}(\sqrt{g}F_{(B)}^{\mu\nu})=0.
\eea
We may find a solution that satisfies the following equations separately
\bea
\frac{1}{\sqrt{g}}\partial_{\rho}(\sqrt{g}F_{(B)}^{\rho\nu})=0\,,\label{eom}\\
\frac{1}{\sqrt{g}}\partial_{a}(\sqrt{g}F_{(B)}^{a\nu})=0\,.
\eea
First of all, let us look at the $\nu=2$ component of (\ref{eom})
\be
\frac{1}{\sqrt{g}}\partial_{\rho}(\sqrt{g}F_{(B)}^{\rho 2})=0
\ee
which is satisfied
if
\be\label{FB}
F^{(B)}_{\rho 2}=\pd_\rho B_2-\partial_2 B_{\rho}=\frac{D_1(x^a)}{(\rho+r_s)^2}\,.
\ee
Supposing $B_{2}$ takes the form
\be
B_2 =-\frac{D_1(x^a)}{\rho+r_s}+D_2(x^a)\,,
\ee
the boundary condition fixes the coefficients $D_1$ and $D_2$
\bea
&& B_2\big{|}_{\rho=\varepsilon}=-\frac{D_1}{r_s+\varepsilon}+D_2=f_2^{(l)}(x^a) \nn\\
&& B_2\big{|}_{\rho=L}=-\frac{D_1}{r_s+L}+D_2=f_2^{(r)}(x^a)\,.
\eea
We can get that
\bea
&& D_1=
\frac{(r_s+\varepsilon)(r_s+L)}{L-\varepsilon}\left(f_2^{(r)}-f_2^{(l)}\right)\,,\nn\\
&& D_2= \frac{r_s+L}{L-\varepsilon}f_2^{(r)}-\frac{r_s+\varepsilon}{L-\varepsilon}f_2^{(l)} \,.
\eea
$\varepsilon$ is much smaller than $r_s$ and $L$, so we can write $r_s\approx r_s+\varepsilon$ and $L\approx L-\varepsilon$. Thus the solution for $B_2$ can be written as
\bea\label{B2}
B_{2}&=& -\frac{r_s(r_s+L)}{(\rho+r_s)L}(f_2^{(r)}-f_2^{(l)})+\frac{r_s+L}{L}f_2^{(r)}-\frac{r_s}{L}f_2^{(l)}\,.
\eea
For a similar reason, we can get the solution for other components of $B_a$, and the solution of $B_a$ can be written as
\bea\label{B3}
B_a &=&-\frac{r_s(r_s+L)}{(\rho+r_s)L}(f_a^{(r)}-f_a^{(l)}) +\frac{r_s+L}{L}f_a^{(r)}-\frac{r_s}{L}f_a^{(l)}\,.
\eea
With the solution $B_{a}$, one can then further fix $B_\rho$ such that the field strength satisfies the bulk equation of motion. We have
\bea
&& \partial_\rho B_a=\frac{1}{(\rho+r_s)^2}\frac{r_s(r_s+L)}{L} (f_a^{(r)}-f_a^{(l)})\nn\\
&& \partial_a B_{\rho}=\frac{1}{(\rho+r_s)^2}\frac{r_s(r_s+L)}{L}(f_a^{(r)}-f_a^{(l)})\,.\nn
\eea
Note that the field strength is set to zero here. Assuming flat boundary configurations
\be
f_a^{(r)}=C_a^{(r)}-\pd_a\lambda^{(r)}\,,~~~~f_a^{(l)}=C_a^{(l)}-\pd_a\lambda^{(l)}\,.
\ee
We can get the solution for $B_{\rho}$ by integrating $\pd_a B_{\rho}$ over $x^a$
\bea\label{Brho}
B_\rho = \frac{1}{(\rho+r_s)^2}\frac{r_s(r_s+L)}{L}\times [x^a\cdot(C_a^{(r)}-C_a^{(l)})+(\lambda^{(r)}-\lambda^{(l)})]\,,
\eea
As a double check, one can put the solutions (\ref{B2}), (\ref{Brho}), and (\ref{B3}) into Eq. (\ref{solBeq}) to check if it is satisfied or not.

Now, we can perform the coordinate transformation and transfer everything into coordinate \eqref{yymetric}. We have
\be
\frac{d\rho}{dx}=\sqrt{\frac{\rho}{\rho+r_s}}\,,
\ee
thus $B_y$ can be written as
\bea
B_y = \frac{\sqrt{\rho}}{(\rho+r_s)^{5/2}}\frac{r_s(r_s+L)}{L} \times [x^a\cdot(C_a^{(r)}-C_a^{(l)})+(\lambda^{(r)}-\lambda^{(l)})]\,.
\eea
So, the solutions can be summarized as
\bea
&& B_a =-\frac{r_s(r_s+L)}{(\rho+r_s)L}[C_a^{(r)}-C_a^{(l)}+\pd_a(\lambda^{(r)}-\lambda^{(l)})]\nn\\
&~& ~~~~~~~+\frac{r_s+L}{L}(C_a^{(r)}+\pd_a\lambda^{(r)})-\frac{r_s}{L}(C_a^{(l)}+\pd_a \lambda^{(l)})\,,\\
&& B_y = \frac{\sqrt{\rho}}{(\rho+r_s)^{5/2}}\frac{r_s(r_s+L)}{L}\times [x^a\cdot(C_a^{(r)}-C_a^{(l)})+(\lambda^{(r)}-\lambda^{(l)})]\,\nn\\
\eea

Then, we need to add those modes in the Euclidean path integral of the U(1) gauge theory. 
Correspondingly, there would be new contributions to the partition function and the thermal entropy. 
Those bulk modes corresponding to the bulk would-be pure gauge configurations. It would be interesting to see the relationship among those modes, the soft hair of black holes \cite{Hawking:2016sgy, He:2014laa, Haco:2018ske, Hawking:2016msc, Strominger2018, Cheng:2020vzw, Cheng:2022xyr, Cheng:2022xgm}, Barnich's nonproper gauge degrees of freedom \cite{Barnich:2018zdg, Barnich:2019qex, Alessio:2020lpk, Aggarwal:2022rrp}, and other boundary modes \cite{Park:2022nkb}.

\section{Effective action for fields $\phi$ and $W$}
\label{appA2}

In this appendix, we derive the effective action (\ref{BHSE}) from the original action of the U(1) gauge theory. The Euclidean action for Maxwell's theory on a curved Euclidean background can be written as
\be
S_E=\frac{1}{4e^2}\int_{\mathcal{M}}d\tau d^3x \sqrt{g} ~F^{\mu\nu}F_{\mu\nu}
\ee
Now we are going to work out the above Euclidean action in terms of fields $\hat A_\mu$, $\phi$, and $W$.  Working on the Euclidean Schwarzschild black hole background (\ref{metricrho}), the action can be separated into two parts with regard to (3+$y$) decomposition as
\bea
S_E &=& \frac{1}{4e^2}\int_\mM d\tau d^3x\sqrt{g}~ F^{ab}F_{ab}
 +\frac{1}{2e^2}\int_\mM d\tau d^3x\sqrt{g}~F^{ya}F_{ya}\,.
\eea
Putting the field decomposition \eqref{decom} into the action, the above effective action can be further written as
\bea
S_E &=& \frac{1}{4e^2}\int_\mM d\tau d^3x\sqrt{g}~ \hat F^{\mu\nu}\hat F_{\mu\nu} +\frac{1}{2e^2 |y|^2}\int_\mM d\tau d^3x \sqrt{g} \left[ g^{ab}\pd_a\phi\pd_b\phi\right]\,\nn\\
&~&-\frac{1}{e^2|y|}\int_\mM d\tau d^3x \sqrt{g} \left[g^{ab}(\pd_y\hat A_a-\pd_a\hat A_y)\pd_b\phi \right]\,.
\eea
Denoting the first part in the above action as $\hat S_0$, the above effective action can be further written as
\bea
S_E &=& \hat S_0  +\frac{1}{2e^2 |y|^2}\int_\mM d\tau d^3x \sqrt{g} \left[ g^{ab}\pd_a\phi\pd_b\phi\right]-\frac{1}{e^2|y|}\int_\mM d\tau d^2x dy ~ \sqrt{g} \left[ g^{ab}(\pd_y\hat A_a-\pd_a\hat A_y)\pd_b\phi \right]\,,\nn\\
\eea
which can be further simplified as
\bea
S_E &=&\hat S_0+\frac{1}{2e^2 |y|^2}\int_\mM d\tau d^3x \sqrt{g} \left[ g^{ab}\pd_a\phi\pd_b\phi\right]
-\frac{i}{e^2|y|}\int d\tau d^2x\left[\sqrt{g} ~ g^{ab}\right]_{y=y_1}^{y=y_2}\pd_a W \pd_b \phi \nn\\
&~&+\frac{1}{e^2|y|}\int_\mM d\tau d^3x~\pd_y(\sqrt{g} g^{ab}) \left[  \hat A_a -\pd_a (\int dy \hat A_y)  \right]\pd_b \phi
\,.\nn
\eea
Note that the last term in the above expression can be set to zero by gauge choice. So we are going to ignore the last term and only focus on the following Euclidean action as shown in Eq. (\ref{BHSE})
\bea
S_{E} &=& \frac{1}{4e^2}\int_\mM d\tau d^3x\sqrt{g}~ \hat F^{\mu\nu}\hat F_{\mu\nu}
+\frac{1}{2e^2 |y|^2}\int_\mM d\tau d^3x \sqrt{g} \left[ g^{ab}\pd_a\phi\pd_b\phi\right]
\nn\\&~&
-\frac{i}{e^2|y|}\int d\tau d^2x\left[\sqrt{g} ~ g^{ab}\right]_{y=y_1}^{y=y_2}\pd_a W \pd_b \phi \,.
\eea

\section{Dimensional reduction}
\label{redim}

This appendix performs the dimensional reduction in Sec. \ref{seceff}, which is to reduce the higher dimensional action $S[\phi,W]$ to three-dimensional action $S_{\phi,W}$ shown below. We have
\bea
S[\phi,W] &=& \frac{1}{2e^2 |y|^2}\int_\mM d\tau d^3x \sqrt{g} \left[ g^{ab}\pd_a\phi\pd_b\phi\right] -\frac{i}{e^2|y|}\int d\tau d^2x\left[\sqrt{g} ~ g^{ab}\right]_{y=y_1}^{y=y_2}\pd_a W \pd_b \phi\,\label{C1}
\eea
and
\bea
S_{\phi,W} &=& \frac{1}{2e'^2}\int d\tau d^2x \sqrt{h} \left[ h^{ab}\pd_a\phi\pd_b\phi\right]\nn\\
&~&-\frac{i}{2e'^2}\int d\tau d^2x \sqrt{h} \left[  \gamma_1~ h^{\tau\tau} \pd_\tau W\pd_\tau \phi +\gamma_2~ h^{\theta\theta} \pd_\theta W\pd_\theta \phi+\gamma_3~ h^{\varphi\varphi} \pd_\varphi W\pd_\varphi \phi\right]\,.\label{C2}
\eea
We are going to divide the problem into two steps. The first is to solve the low dimensional metric $h_{ab}$ and coupling constant $e'$. Then we are able to solve $\gamma_1$, $\gamma_2$, and $\gamma_3$ with the results from the first step.

\subsection{Metric and coupling constant $e'$}

As discussed in the main context, we assume the three-dimensional metric takes the following form
\be
h_{ab}=\text{diag}(h_{\tau\tau},~R^2,~R^2\sin^2\theta)\,,
\ee
with topology $ {S}^1\times {S}^2$. The radius of $ {S}^1$ is $\beta$ and radius of ${S}^2$ is $R$. We just need to solve the following equations
\bea
\frac{1}{2e^2 |y|^2}\int d\tau d^2x ~(\int_{y_1}^{y_2} dy \sqrt{g}g^{ab}) ~\pd_a\phi\pd_b\phi = \frac{1}{2e'^2}\int d\tau d^2x \sqrt{h} \left[ h^{ab}\pd_a\phi\pd_b\phi\right]\,.\nn
\eea
The above equations can be simplified as
\be
\frac{1}{2e^2 |y|^2}\int_{\varepsilon}^{L} d\rho~\frac{\partial y}{\pd \rho} \sqrt{g}g^{ab}=\frac{1}{2e'^2}\sqrt{h}h^{ab}\,.
\ee
We are left with two independent components
\bea
&& \frac{1}{e^2 |y|^2}(3r_s^2L+\frac{3}{2}r_s L^2+\frac{L^3}{3}+r_s^3\ln\frac{L}{\varepsilon}) = \frac{1}{e'^2}\sqrt{h^{\tau\tau}}R^2\,,\\
&& \frac{L}{e^2 |y|^2} = \frac{1}{e'^2}\sqrt{h_{\tau\tau}}\,.
\eea
There are three unknown variables and only two independent equations. So we are going to write $h_{\tau\tau}$ and coupling constant $e'^2$ as a function of radius $R$. The solution can be written as
\bea
h_{\tau\tau} &=& \frac{L R^2}{3r_s^2L+\frac{3}{2}r_s L^2+\frac{L^3}{3}+r_s^3\ln{L}/{\varepsilon}}\,,\\
\frac{1}{e'^2} &=& \frac{L}{e^2|y|^2}\sqrt{h^{\tau\tau}}=\frac{\sqrt{L}}{e^2|y|^2}\frac{\sqrt{3r_s^2L+\frac{3}{2}r_s L^2+\frac{L^3}{3}+r_s^3\ln{L}/{\varepsilon}}}{ R}\,.\nn\\
\eea

\subsection{$\gamma$ couplings}

The next step is to calculate $\gamma_1$, $\gamma_2$, and $\gamma_3$. To do that we just need to match the rest of the actions (\ref{C1}) and (\ref{C2}), which read as
\bea
& \frac{1}{e^2|y|}\int d\tau d^2x\left[\sqrt{g} ~ g^{ab}\right]_{y=y_1}^{y=y_2}\pd_a W \pd_b \phi=~~~~~~~~~~~~~~~~~~~~~~~~~ \nn\\
& \frac{1}{2e'^2}\int d\tau d^2x \sqrt{h} \left[  \gamma_1~ h^{\tau\tau} \pd_\tau W\pd_\tau \phi +\gamma_2~ h^{\theta\theta} \pd_\theta W\pd_\theta \phi+\gamma_3~ h^{\varphi\varphi} \pd_\varphi W\pd_\varphi \phi\right]\,.\nn\\
\eea
The useful information from the above equation is
\bea
\frac{1}{e^2|y|}\left[\sqrt{g} ~ g^{\tau\tau}\right]_{y=y_1}^{y=y_2} &=& \frac{\gamma_1}{2e'^2}\sqrt{h} ~ h^{\tau\tau}\,,\\
\frac{1}{e^2|y|}\left[\sqrt{g} ~ g^{\theta\theta}\right]_{y=y_1}^{y=y_2} &=& \frac{\gamma_2}{2e'^2}\sqrt{h} ~ h^{\theta\theta} \,,\\
\gamma_2 &=& \gamma_3\,.
\eea
Writing everything explicitly, we have
\bea
\frac{1}{e^2|y|}\left[ \sqrt{\frac{\rho(y)+r_s}{\rho}}(\rho(y)+r_s)^2\right]_{y=y_1}^{y=y_2} &=& \gamma_1~\frac{L}{2e^2|y|^2}h^{\tau\tau}R^2\,,\\
\frac{1}{e^2|y|}\left[\sqrt{\frac{\rho(y)}{\rho(y)+r_s}}\right]_{y=y_1}^{y=y_2} &=& \gamma_2~\frac{L}{2e^2|y|^2}\,,\\
\gamma_2 &=& \gamma_3\,.
\eea
The above equations can be solved as
\bea
\gamma_1 &=& -2|y|\sqrt{\frac{r_s}{\varepsilon}}\frac{r_s^2 }{3r_s^2L+\frac{3}{2}r_s L^2+\frac{L^3}{3}+r_s^3\ln{L}/{\varepsilon}}\,\nn\\
\gamma_2 &=&\gamma_3 = \frac{2|y|}{L}\sqrt{\frac{L}{L+r_s}} \,.
\eea



\section{Bulk partition function}
\label{Sch}

In this appendix, we give a detailed analysis of the partition function of the bulk fluctuation modes.
The gauge fixing condition can be imposed by inserting the following identity in the path integral.

\be
1=\int \mD \lambda \det \left(\frac{\pd G}{\pd \lambda}\right)\delta(G-0)\,,\label{gaugefix}
\ee
with gauge fixing condition $G$. 
After gauge fixing, there are only two polarization degrees of freedom for Maxwell's theory in the bulk, which can be simulated by two massless scalar fields.
One can repeat the calculation using the standard Faddeev-Popov method. Here we would just assume that on black hole background, the gauge fields $A_{\mu}$ are also left with two massless bosonic components after gauge fixing.

The metric of Schwarzschild is shown in (\ref{metric}). For one free massless particle living on this background, the motion for geodesics can be expressed as
\be\label{geo}
g_{\mu\nu}\frac{dx^\mu }{d\lambda}\frac{dx^\nu}{d\lambda}=0\,,
\ee
with $\lambda$ the parameter along the trajectory.
For a Schwarzschild black hole (See Sec. 5.4 in \cite{carrollSpacetimeGeometry2019} for more details), the Killing vector associated with energy can be written as
\be
K^{\mu}=(\pd_\tau)^{\mu}=(1,0,0,0)\,~~~~\Bigg(\text{or }K_{\mu}=\Big(1-\frac{2G_N M}{r},0,0,0\Big)\Bigg)\,.
\ee
And the Killing vector associated with angular momentum is
\be
R^{\mu}=(\pd_{\varphi})^{\mu}=(0,0,0,1)\,~~~~(\text{or }R_\mu =(0,0,0,r^2\sin^2\theta))\,.
\ee
There are two conserved charges energy and angular momentum on the equatorial plane because of the Killing vectors. The conserved quantities can be expressed as
\bea
 \textbf{E} &=& -K_{\mu}\frac{dx^\mu}{d\lambda}=-\left(1-\frac{2G_N M}{r}\right)\frac{d\tau}{d\lambda}\nn\\
 \textbf{L} &=& R_{\mu}\frac{dx^{\mu}}{d\lambda}=r^2\frac{d\varphi}{d\lambda}\,.
\eea
On the other hand, we have
\be
p_\mu=g_{\mu\nu}\frac{dx^\nu}{d\lambda}\,.
\ee
Then expression (\ref{geo}) can be expressed by the conserved energy and angular momentum as
\be
-\left(1- \frac{2G_N M}{r}\right)^{-1} \textbf{E}^2+\left(1- \frac{2G_N M}{r}\right) p_r^2+\frac{ \textbf{L}^2}{r^2}=0\,,
\ee
which can be rewritten as
\be\label{pr}
p_r^2 = \left(1- \frac{2G_N M}{r}\right)^{-1}\left[\left(1- \frac{2G_N M}{r}\right)^{-1}  \textbf{E}^2+\frac{  \textbf{L}^2}{r^2}\right]\,.
\ee
For a massless scalar field with mode expansion
\be
\bar A(\tau,r,\theta,\varphi)=\sum_{\omega}\sum_{l,m}e^{-i\omega\tau}Y_{lm}(\theta,\phi) \tilde A(\omega, l, m, r)\,,
\ee
with
\be
\pd_r \tilde A=ip_r\tilde A\,,
\ee
the above expression (\ref{pr}) can be written as the dispersion relation for $A$
\be
p_r^2=\left(1- \frac{2G_N M}{r}\right)^{-1}\left[\left(1- \frac{2G_N M}{r}\right)^{-1}\omega^2+\frac{l(l+1)}{r^2}\right]\,.
\ee
This expression can also be obtained from different methods, for example from the equation of motion of massless field on curved space time \cite{tHooft:1984kcu}.

Let us study the statistical properties of those bulk fluctuation modes. Because of the boundary conditions on $r=r_s+\varepsilon$ and $r=r_s+L$, we have a standing-wave condition along the radius direction, which can be written as
\be
n\pi= \int_{r_s+\varepsilon}^{r_s+L} p_r (r,\omega,l)~ dr\,,
\ee
with $n\in  {Z}_n$. The partition function of bulk fluctuation modes can always be written as
\be
\ln Z_{\bar A}=-2 \sum_{\omega} \ln (1-e^{-\beta \omega})\,,
\ee
where the factor 2 means we have two polarizations, and we ignored the zero-point energy in the above expression.
Now we can change the summation of $\omega$ into integration by introducing density of state $g(\omega)$ and regarding the spectrum to be continuous. We obtain
\bea
\ln Z_F &=& -2\int_0^\infty g(\omega)\ln(1-e^{-\beta \omega})d\omega
= -2\int_0^\infty \ln(1-e^{-\beta \omega})d\Gamma(\omega)\nn\\
&=& -2\ln(1-e^{-\beta \omega})\Gamma(\omega)\big{|}_{0}^{\infty} +2\int_0^\infty \frac{\Gamma(\omega)e^{-\beta \omega}}{1-e^{-\beta \omega}}\beta d\omega\,,
\eea
where $\Gamma(\omega)$ defined by $d\Gamma =g(\omega)d\omega$ is the number of state not exceeding $\omega$. The first part is zero when $\omega\to 0$ and $\omega\to \infty$, so the final result for our partition function can be written as
\be\label{ZFA}
\ln Z_F =2\beta\int_0^\infty\frac{\Gamma(\omega)}{e^{\beta \omega}-1}d\omega\,.
\ee
$\Gamma(\omega)$ is the number of states that have energy lower than $\omega$, and we have
\bea
&\Gamma(\omega)
= \frac{1}{\pi}\sum_{l} (2l+1)\int_{r_s+\varepsilon}^{r_s+L}dr \sqrt{\left(1- \frac{2G_N M}{r}\right)^{-1}\left[\left(1- \frac{2G_N M}{r}\right)^{-1}\omega^2+\frac{l(l+1)}{r^2}\right]}\nn\\
&\approx  \frac{1}{\pi}\int_l (2l+1)dl \int_{r_s+\varepsilon}^{r_s+L}\frac{dr}{1- \frac{2G_N M}{r}}\left[\omega^2+\left(1- \frac{2G_N M}{r}\right)\frac{l(l+1)}{r^2}\right]^{1/2}\,.~~~~~~~~~~~\nn
\eea
\be
~\label{Gamma}
\ee
Note that we have changed the summation of $l$ into an integral assuming the area of the boundary is big enough. The summation or integration is from $l=0$ to the state with energy $\omega$. Now we can put (\ref{Gamma}) into Eq. (\ref{ZFA}), and the logarithm of the partition function can be written as
\bea
\ln Z_F 
= \frac{2\beta}{\pi}\int_0^\infty\frac{d\omega}{e^{\beta \omega}-1}\int d[l(l+1)] \int~\frac{dr}{1- \frac{2G_N M}{r}}\left[\omega^2+\left(1- \frac{2G_N M}{r}\right)\frac{l(l+1)}{r^2}\right]^{1/2}\,.\nn
\eea
Now, let us redefine $x=\left(1- \frac{2G_N M}{r}\right)\frac{l(l+1)}{r^2}$. We obtain
\be
d[l(l+1)]=\frac{r^2}{1- \frac{2G_N M}{r}}dx\,,
\ee
thus we can rewrite the integral as
\bea
\ln Z_F &=& \frac{2\beta}{\pi}\int_0^\infty\frac{d\omega}{e^{\beta \omega}-1} \int~\frac{r^2 dr}{(1- \frac{2G_N M}{r})^2}\int_0^{\omega^2} dx\left[\omega^2+x \right]^{1/2}\nn\\
&=& -\frac{4\beta}{3\pi}\int_0^\infty\frac{\omega^3d\omega}{e^{\beta \omega}-1} \int~\frac{r^4 dr}{(r-2G_N M)^2}\,.
\eea
Those integrals are straightforward to work out, we have
\be
\int_0^\infty\frac{\omega^3d\omega}{e^{\beta \omega}-1}=\frac{\pi^4}{15}\frac{1}{\beta^4}
\ee
and
\bea
\int_{r_s+\varepsilon}^{r_s+l}~\frac{r^4 dr}{(r-2G_N M)^2}
\approx   \frac{r_s^4}{\varepsilon} +4~r_s^3\ln \frac{L}{\varepsilon}-\frac{r_s^4}{L}+6~r_s^2 L+2~r_s L^2+\frac{L^3}{3}\,.\nn\\
\eea
All in all, the logarithm of the partition function can be written as
\bea\label{partbulk}
\ln Z_F 
&=& -\frac{4\pi^3}{45}\frac{1}{\beta^3}\frac{r_s^4}{\varepsilon} -\frac{16\pi^3}{45}\frac{r_s^3}{\beta^3}\ln \frac{L}{\varepsilon}
-\frac{4\pi^3}{45}\frac{1}{\beta^3}\left( -\frac{r_s^4}{L}+6~r_s^2 L+2~r_s L^2+\frac{L^3}{3}\right)\,.
\eea
The corresponding entropy $\mS_F = (1-\beta\pd_\beta)$ can be calculated as
\bea\label{bulkentropy}
\mS_F 
 =\frac{16\pi^3}{45}\frac{1}{\beta^3}\frac{r_s^4}{\varepsilon}+\frac{64\pi^3}{45}\frac{r_s^3}{\beta^3}\ln \frac{L}{\varepsilon}+\frac{16\pi^3}{45}\frac{1}{\beta^3}\left( -\frac{r_s^4}{L}+6~r_s^2 L+2~r_s L^2+\frac{L^3}{3}\right)\,.
\eea


\providecommand{\href}[2]{#2}\begingroup\raggedright\endgroup


\begin{thebibliography}{100}

\bibitem{Hawking1971}
S.~W. Hawking, ``Gravitational radiation from colliding black holes,''
  \href{http://dx.doi.org/10.1103/physrevlett.26.1344}{{\em Phys. Rev. Lett.}
  {\bfseries 26} (1971) 1344--1346}.

\bibitem{Christodoulou1970}
D.~Christodoulou, ``Reversible and irreversible transformations in black-hole
  physics,'' \href{http://dx.doi.org/10.1103/physrevlett.25.1596}{{\em Phys.
  Rev. Lett.} {\bfseries 25} (1970) 1596--1597}.

\bibitem{Christodoulou1971}
D.~Christodoulou and R.~Ruffini, ``Reversible transformations of a charged
  black hole,'' \href{http://dx.doi.org/10.1103/physrevd.4.3552}{{\em Phys.
  Rev. D} {\bfseries 4} (1971) 3552--3555}.

\bibitem{HAWKING1974}
S.~W. Hawking, ``Black hole explosions?,''
  \href{http://dx.doi.org/10.1038/248030a0}{{\em Nature} {\bfseries 248} (1974) 30--31}.

\bibitem{Bekenstein1973}
J.~D. Bekenstein, ``Black holes and entropy,''
  \href{http://dx.doi.org/10.1103/physrevd.7.2333}{{\em Phys. Rev. D}
  {\bfseries 7} (1973) 2333--2346}.

\bibitem{Bardeen1973}
J.~M. Bardeen, B.~Carter, and S.~W. Hawking, ``The four laws of black hole
  mechanics,'' \href{http://dx.doi.org/10.1007/bf01645742}{{\em Comm. Math.
  Phys.} {\bfseries 31} (1973) 161--170}.

\bibitem{Bekenstein1972}
J.~D. Bekenstein, ``Black holes and the second law,''
  \href{http://dx.doi.org/10.1007/bf02757029}{{\em Lettere Al Nuovo Cimento
  Series 2} {\bfseries 4} (1972) 737--740}.

\bibitem{CARTER1972}
B.~Carter, ``Rigidity of a black hole,''
  \href{http://dx.doi.org/10.1038/physci238071b0}{{\em Nature Physical Science}
  {\bfseries 238} (1972) 71--72}.

\bibitem{Almheiri2020}
A.~Almheiri, T.~Hartman, J.~Maldacena, E.~Shaghoulian, and A.~Tajdini, ``The
  entropy of {{Hawking}} radiation,''
  \href{http://10.1103/RevModPhys.93.035002}{{\em Rev. Mod. Phys.}
  {\bfseries 93} (2021) 035002},
  \href{http://arxiv.org/abs/2006.06872v1}{{\ttfamily arXiv:2006.06872v1}}.

\bibitem{tHooft:1984kcu}
G.~'t~Hooft, ``On the quantum structure of a black hole,''
  \href{http://dx.doi.org/10.1016/0550-3213(85)90418-3}{{\em Nuclear Phys. B
  Proc. Suppl.} {\bfseries 256} (1985) 727--745}.

\bibitem{Srednicki1993}
M.~Srednicki, ``Entropy and area,''
  \href{http://dx.doi.org/10.1103/physrevlett.71.666}{{\em Phys. Rev. Lett.}
  {\bfseries 71} (1993) 666--669}.

\bibitem{Susskind1994}
L.~Susskind and J.~Uglum, ``Black hole entropy in canonical quantum gravity and
  superstring theory,'' \href{http://dx.doi.org/10.1103/PhysRevD.50.2700}{{\em
  Phys Rev D} {\bfseries 50} (1994) 2700--2711},
  \href{http://arxiv.org/abs/hep-th/9401070}{{\ttfamily arXiv:hep-th/9401070
  [hep-th]}}.

\bibitem{Solodukhin2011}
S.~N. Solodukhin, ``Entanglement entropy of black holes,''
  \href{http://dx.doi.org/10.12942/lrr-2011-8}{{\em Living Rev Relativ}
  {\bfseries 14} (2011) 8},
  \href{http://arxiv.org/abs/1104.3712}{{\ttfamily arXiv:1104.3712 [hep-th]}}.

\bibitem{Blommaert2018a}
A.~Blommaert, T.~G. Mertens, and H.~Verschelde, ``Edge dynamics from the path
  integral: {Maxwell and Yang-Mills},''
  \href{http://dx.doi.org/10.1007/JHEP11(2018)080}{{\em JHEP}
  {\bfseries 11} (2018) 080},
  \href{http://arxiv.org/abs/1804.07585v2}{{\ttfamily arXiv:1804.07585v2
  [hep-th]}}.

\bibitem{Donnelly:2014fua}
W.~Donnelly and A.~C. Wall, ``Entanglement entropy of electromagnetic edge
  modes,'' \href{http://dx.doi.org/10.1103/PhysRevLett.114.111603}{{\em Phys.
  Rev. Lett.} {\bfseries 114} (2014) 111603},
  \href{http://arxiv.org/abs/1412.1895v2}{{\ttfamily arXiv:1412.1895v2}}.

\bibitem{Donnelly2015a}
W.~Donnelly and A.~C.~Wall,
``Geometric entropy and edge modes of the electromagnetic field,''
\href{http://10.1103/PhysRevD.94.104053}{{\em Phys. Rev. D} {\bfseries 94} (104053) },
  \href{http://arxiv.org/abs/1506.05792}{{\ttfamily arXiv:1506.05792
  [hep-th]}}.


\bibitem{Wei2015am}
S.~W.~Wei and Y.~X.~Liu,
``Insight into the Microscopic Structure of an AdS Black Hole from a Thermodynamical Phase Transition,''
\href{http://10.1103/PhysRevLett.115.111302}{{\em Phys. Rev. Lett.} {\bfseries 115} (2015) 111302 },
  \href{http://arxiv.org/abs/1502.00386}{{\ttfamily arXiv:1502.00386 [gr-qc]}}.



\bibitem{Haco2018}
S.~Haco, S.~W. Hawking, M.~J. Perry, and A.~Strominger, ``Black hole entropy
  and soft hair,'' \href{http://dx.doi.org/10.1007/JHEP12(2018)098}{{\em JHEP} {\bfseries 12}  (2018) 098},
  \href{http://arxiv.org/abs/1810.01847}{{\ttfamily arXiv:1810.01847
  [hep-th]}}.

\bibitem{Krasnov1996}
K.~V. Krasnov, ``Counting surface states in the loop quantum gravity,''
  \href{http://dx.doi.org/10.1103/PhysRevD.55.3505}{{\em Phys.Rev. D} {\bfseries 55} (1997)
  3505-3513},
  \href{http://arxiv.org/abs/gr-qc/9603025}{{\ttfamily arXiv:gr-qc/9603025
  [gr-qc]}}.

\bibitem{Rovelli1996}
C.~Rovelli, ``Black hole entropy from loop quantum gravity,''
  \href{http://dx.doi.org/10.1103/PhysRevLett.77.3288}{{\em
  Phys. Rev. Lett.} {\bfseries 77} (1996)
  3288--3291}, \href{http://arxiv.org/abs/gr-qc/9603063}{{\ttfamily
  arXiv:gr-qc/9603063 [gr-qc]}}.

\bibitem{Ashtekar1997}
A.~Ashtekar, J.~Baez, A.~Corichi, and K.~Krasnov, ``Quantum geometry and black
  hole entropy,'' \href{http://dx.doi.org/10.1103/PhysRevLett.80.904}{{\em
  Phys. Rev. Lett. } {\bfseries 80} (1998) 904-907},
  \href{http://arxiv.org/abs/gr-qc/9710007}{{\ttfamily arXiv:gr-qc/9710007
  [gr-qc]}}.

\bibitem{Domagala2004}
M.~Domagala and J.~Lewandowski, ``Black hole entropy from quantum geometry,''
  \href{http://dx.doi.org/10.1088/0264-9381/21/22/014}{{\em Class. Quant. Grav.} {\bfseries 21} (2004) 5233-5244},
  \href{http://arxiv.org/abs/gr-qc/0407051}{{\ttfamily arXiv:gr-qc/0407051
  [gr-qc]}}.

\bibitem{Meissner2004}
K.~A. Meissner, ``Black hole entropy in loop quantum gravity,''
  \href{http://dx.doi.org/10.1088/0264-9381/21/22/015}{{\em Class. Quant. Grav.
  } {\bfseries 21} (2004) 5245-5252},
  \href{http://arxiv.org/abs/gr-qc/0407052}{{\ttfamily arXiv:gr-qc/0407052
  [gr-qc]}}.

\bibitem{Ghosh2006}
A.~Ghosh and P.~Mitra, ``Counting black hole microscopic states in loop quantum
  gravity,'' \href{http://dx.doi.org/10.1103/physrevd.74.064026}{{\em Phys.
  Rev. D} {\bfseries 74} (2006) 064026}.

\bibitem{Cardy1986}
J.~L. Cardy, ``Operator content of two-dimensional conformally invariant
  theories,'' \href{http://dx.doi.org/10.1016/0550-3213(86)90552-3}{{\em
  Nuclear Phys. B Proc. Suppl.} {\bfseries 270} (1986) 186--204}.

\bibitem{Bloete1986}
H.~W.~J. Blöte, J.~L. Cardy, and M.~P. Nightingale, ``Conformal invariance,
  the central charge, and universal finite-size amplitudes at criticality,''
  \href{http://dx.doi.org/10.1103/physrevlett.56.742}{{\em Phys. Rev. Lett.}
  {\bfseries 56} (1986) 742--745}.

\bibitem{Strominger1998}
A.~Strominger, ``Black hole entropy from near-horizon microstates,''
  \href{http://dx.doi.org/10.1088/1126-6708/1998/02/009}{{\em JHEP} {\bfseries 02} (1998) 009--009},
  \href{http://arxiv.org/abs/hep-th/9712251}{{\ttfamily arXiv:hep-th/9712251
  [hep-th]}}.

\bibitem{Birmingham1998}
D.~Birmingham, I.~Sachs, and S.~Sen, ``Entropy of three-dimensional black holes
  in string theory,''
  \href{http://dx.doi.org/10.1016/s0370-2693(98)00236-6}{{\em Phys. Lett. B}
  {\bfseries 424} (1998) 275--280},
  \href{http://arxiv.org/abs/hep-th/9801019}{{\ttfamily arXiv:hep-th/9801019
  [hep-th]}}.

\bibitem{Guica2009}
M.~Guica, T.~Hartman, W.~Song, and A.~Strominger, ``The {Kerr/CFT}
  correspondence,'' \href{http://dx.doi.org/10.1103/physrevd.80.124008}{{\em
  Phys. Rev. D} {\bfseries 80} (2009) 124008}.

\bibitem{Carlip2011}
S.~Carlip, ``Effective conformal descriptions of black hole entropy,''
  \href{http://dx.doi.org/10.3390/e13071355}{{\em Entropy } {\bfseries 13} (2011)
  1355-1379}, \href{http://arxiv.org/abs/1107.2678}{{\ttfamily
  arXiv:1107.2678 [gr-qc]}}.

\bibitem{Brown1986}
J.~D. Brown and M.~Henneaux, ``{Central Charges in the Canonical Realization of
  Asymptotic Symmetries: An Example from Three-Dimensional Gravity},''
  \href{http://dx.doi.org/10.1007/BF01211590}{{\em Commun. Math. Phys.}
  {\bfseries 104} (1986) 207--226}.

\bibitem{Solodukhin1999}
S.~N. Solodukhin, ``Conformal description of horizon's states,''
  \href{http://dx.doi.org/10.1016/S0370-2693(99)00398-6}{{\em Phys. Lett. B}
  {\bfseries 454} (1999) 213--222},
  \href{http://arxiv.org/abs/hep-th/9812056}{{\ttfamily arXiv:hep-th/9812056
  [hep-th]}}.

\bibitem{Carlip1999a}
S.~Carlip, ``Black hole entropy from conformal field theory in any dimension,''
  \href{http://dx.doi.org/10.1103/PhysRevLett.82.2828}{{\em Phys. Rev. Lett.}
  {\bfseries 82} (1999) 2828--2831},
  \href{http://arxiv.org/abs/hep-th/9812013}{{\ttfamily arXiv:hep-th/9812013
  [hep-th]}}.

\bibitem{Carlip1999}
S.~Carlip, ``Entropy from conformal field theory at Killing horizons,''
  \href{http://dx.doi.org/10.1088/0264-9381/16/10/322}{{\em
  Class.Quant.Grav.} {\bfseries 16} (1999) 3327-3348 },
  \href{http://arxiv.org/abs/gr-qc/9906126}{{\ttfamily arXiv:gr-qc/9906126
  [gr-qc]}}.

\bibitem{Silva2002}
S.~Silva, ``Black hole entropy and thermodynamics from symmetries,''
  \href{http://dx.doi.org/10.1088/0264-9381/19/15/306}{{\em Class. Quant. Grav.}
  {\bfseries 19} (2002) 3947-3962},
  \href{http://arxiv.org/abs/hep-th/0204179}{{\ttfamily arXiv:hep-th/0204179
  [hep-th]}}.

\bibitem{Carlip2007}
S.~Carlip, ``Symmetries, horizons, and black hole entropy,''
  \href{http://dx.doi.org/10.1007/s10714-007-0467-6}{{\em
  Gen. Rel. Grav.} {\bfseries 39} (2007) 1519-1523
  }, \href{http://arxiv.org/abs/0705.3024}{{\ttfamily arXiv:0705.3024
  [gr-qc]}}.

\bibitem{Castro2010}
A.~Castro, A.~Maloney, and A.~Strominger, ``Hidden conformal symmetry of the
  Kerr black hole,'' \href{http://dx.doi.org/10.1103/PhysRevD.82.024008}{{\em
  Phys. Rev. D} {\bfseries 82} (2010) 024008 },
  \href{http://arxiv.org/abs/1004.0996}{{\ttfamily arXiv:1004.0996 [hep-th]}}.

\bibitem{Wang2010}
Y.-Q. Wang and Y.-X. Liu, ``Hidden conformal symmetry of the {Kerr-Newman} black
  hole,'' \href{http://dx.doi.org/10.1007/jhep08(2010)087}{{\em JHEP} {\bfseries 8} (2010) 087}.

\bibitem{Chen2010}
C.-M. Chen and J.-R. Sun, ``Hidden conformal symmetry of the
  {Reissner-Nordstr\"om} black holes,''
  \href{http://dx.doi.org/10.1007/jhep08(2010)034}{{\em JHEP}
  {\bfseries 08} (2010) 034}.

\bibitem{Carlip2011a}
S.~Carlip, ``Extremal and nonextremal {Kerr/CFT} correspondences,''
  \href{http://dx.doi.org/10.1007/jhep04(2011)076}{{\em JHEP}
  {\bfseries 04} (2011) 076 }.

\bibitem{Compere2017}
G.~Comp{\`{e}}re, ``The {Kerr/CFT} correspondence and its extensions,''
  \href{http://dx.doi.org/10.1007/s41114-017-0003-2}{{\em Living Rev. Relativ.}
  {\bfseries 20} (2017) 1}.

\bibitem{Song2012}
W.~Song and A.~Strominger, ``Warped {AdS}$_3$/dipole-{CFT} duality,''
  \href{http://dx.doi.org/10.1007/jhep05(2012)120}{{\em JHEP}
  {\bfseries 05} (2012) 120}.

\bibitem{Carlip2018}
S.~Carlip, ``Black hole entropy from {BMS} symmetry at the horizon,''
  \href{http://dx.doi.org/10.1103/PhysRevLett.120.101301}{{\em Phys. Rev.
  Lett.} {\bfseries 120} (2018) 101301 },
  \href{http://arxiv.org/abs/1702.04439}{{\ttfamily arXiv:1702.04439 [gr-qc]}}.

\bibitem{Carlip2019}
S.~Carlip, ``Near-horizon {BMS} symmetry, dimensional reduction, and black hole
  entropy,'' \href{http://dx.doi.org/10.1103/PhysRevD.101.046002}{{\em Phys.
  Rev. D } {\bfseries 101} (2020) 046002},
  \href{http://arxiv.org/abs/1910.01762}{{\ttfamily arXiv:1910.01762
  [hep-th]}}.

\bibitem{Strominger1996}
A.~Strominger and C.~Vafa, ``Microscopic origin of the {Bekenstein-Hawking}
  entropy,'' \href{http://dx.doi.org/10.1016/0370-2693(96)00345-0}{{\em
  Phys. Lett. B} {\bfseries 379} (1996) 99--104},
  \href{http://arxiv.org/abs/hep-th/9601029}{{\ttfamily arXiv:hep-th/9601029
  [hep-th]}}.

\bibitem{Callan1996}
C.~G. Callan and J.~M. Maldacena, ``D-brane approach to black hole quantum
  mechanics,'' \href{http://dx.doi.org/10.1016/0550-3213(96)00225-8}{{\em
  Nuclear Phys. B Proc. Suppl.} {\bfseries 472}(1996) 591--608},
  \href{http://arxiv.org/abs/hep-th/9602043}{{\ttfamily arXiv:hep-th/9602043
  [hep-th]}}.

\bibitem{Peet2000}
A.~W. Peet, ``{TASI} lectures on black holes in string theory,''
  \href{http://arxiv.org/abs/hep-th/0008241}{{\ttfamily arXiv:hep-th/0008241
  [hep-th]}}.

\bibitem{Das2001}
S.~R. Das and S.~D. Mathur, ``The quantum physics of black holes: Results from
  string theory,'' \href{http://dx.doi.org/10.1146/annurev.nucl.50.1.153}{{\em
  Ann. Rev. Nucl. Part. Sci.} {\bfseries 50} (2000) 153-206},
  \href{http://arxiv.org/abs/gr-qc/0105063}{{\ttfamily arXiv:gr-qc/0105063
  [gr-qc]}}.

\bibitem{Maldacena1996}
J.~Maldacena and A.~Strominger, ``Black hole greybody factors and {D-brane}
  spectroscopy,'' \href{http://dx.doi.org/10.1103/PhysRevD.55.861}{{\em
  Phys. Rev. D} {\bfseries 55} (1996) 861--870},
  \href{http://arxiv.org/abs/hep-th/9609026}{{\ttfamily arXiv:hep-th/9609026}}.

\bibitem{Emparan2006}
R.~Emparan, ``Black hole entropy as entanglement entropy: a holographic
  derivation,'' \href{http://dx.doi.org/10.1088/1126-6708/2006/06/012}{{\em
  JHEP} {\bfseries 06} (2006) 012},
  \href{http://arxiv.org/abs/hep-th/0603081}{{\ttfamily arXiv:hep-th/0603081
  [hep-th]}}.

\bibitem{Cheng2022}
P.~Cheng, ``Gauge theory with nontrivial boundary conditions,''
  \href{http://arxiv.org/abs/2302.03233}{{\ttfamily arXiv:2302.03233 [hep-th]}}.

\bibitem{Amsel:2009ev}
A.~J.~Amsel, G.~T.~Horowitz, D.~Marolf and M.~M.~Roberts,
``No Dynamics in the Extremal Kerr Throat,''
\href{http://dx.doi.org/10.1088/1126-6708/2009/09/044}{{\em
  JHEP} {\bfseries 09} (2009) 044},
  \href{http://arxiv.org/abs/0906.2376}{{\ttfamily arXiv:0906.2376 [hep-th]}}.


\bibitem{Dias:2009ex}
O.~J.~C.~Dias, H.~S.~Reall and J.~E.~Santos,
``Kerr-CFT and gravitational perturbations,''
\href{http://dx.doi.org/10.1088/1126-6708/2009/08/101}{{\em
  JHEP} {\bfseries 08} (2009) 101},
  \href{http://arxiv.org/abs/0906.2380}{{\ttfamily arXiv:0906.2380 [hep-th]}}.

\bibitem{Johnstone:2013ioa}
M.~Johnstone, M.~M.~Sheikh-Jabbari, J.~Simon and H.~Yavartanoo,
``Extremal black holes and the first law of thermodynamics,''
\href{http://dx.doi.org/10.1103/PhysRevD.88.101503}{Phys. Rev. D \textbf{88} (2013), 101503}.

\bibitem{Hajian:2013lna}
K.~Hajian, A.~Seraj and M.~M.~Sheikh-Jabbari,
``NHEG Mechanics: Laws of Near Horizon Extremal Geometry (Thermo)Dynamics,''
\href{http://dx.doi.org/10.1007/JHEP03(2014)014}{JHEP \textbf{03} (2014), 014}.

\bibitem{Hajian:2014twa}
K.~Hajian, A.~Seraj and M.~M.~Sheikh-Jabbari,
``Near Horizon Extremal Geometry Perturbations: Dynamical Field Perturbations vs. Parametric Variations,''
\href{http://dx.doi.org/10.1007/JHEP10(2014)111}{JHEP \textbf{10} (2014), 111}.

\bibitem{Susskind:1993if}
L.~Susskind, L.~Thorlacius, and J.~Uglum, ``The stretched horizon and black
  hole complementarity,''
  \href{http://dx.doi.org/10.1103/physrevd.48.3743}{{\em Phys. Rev. D}
  {\bfseries 48} (1993) 3743--3761}.

%

\bibitem{Maldacena:1996ds}
J.~M. Maldacena and L.~Susskind, ``{D-branes and fat black holes},''
  \href{http://dx.doi.org/10.1016/0550-3213(96)00323-9}{{\em Nucl. Phys. B}
  {\bfseries 475} (1996) 679--690},
  \href{http://arxiv.org/abs/hep-th/9604042}{{\ttfamily arXiv:hep-th/9604042}}.

\bibitem{Maldacena:1997ih}
J.~M. Maldacena and A.~Strominger, ``{Universal low-energy dynamics for
  rotating black holes},''
  \href{http://dx.doi.org/10.1103/PhysRevD.56.4975}{{\em Phys. Rev. D}
  {\bfseries 56} (1997) 4975--4983},
  \href{http://arxiv.org/abs/hep-th/9702015}{{\ttfamily arXiv:hep-th/9702015}}.

\bibitem{Banerjee:2010qc}
S.~Banerjee, R.~K. Gupta, and A.~Sen, ``{Logarithmic Corrections to Extremal
  Black Hole Entropy from Quantum Entropy Function},''
  \href{http://dx.doi.org/10.1007/JHEP03(2011)147}{{\em JHEP} {\bfseries 03}
  (2011) 147}, \href{http://arxiv.org/abs/1005.3044}{{\ttfamily arXiv:1005.3044
  [hep-th]}}.

\bibitem{Sen:2012cj}
A.~Sen, ``{Logarithmic Corrections to Rotating Extremal Black Hole Entropy in
  Four and Five Dimensions},''
  \href{http://dx.doi.org/10.1007/s10714-012-1373-0}{{\em Gen. Rel. Grav.}
  {\bfseries 44} (2012) 1947--1991},
  \href{http://arxiv.org/abs/1109.3706}{{\ttfamily arXiv:1109.3706 [hep-th]}}.

\bibitem{Iliesiu:2020qvm}
L.~V. Iliesiu and G.~J. Turiaci, ``The statistical mechanics of near-extremal
  black holes,'' 
  \href{http://dx.doi.org/10.1007/JHEP05(2021)145}{{\em JHEP} {\bfseries 05}
  (2011) 145},
  \href{http://arxiv.org/abs/2003.02860}{{\ttfamily
  arXiv:2003.02860}}.

\bibitem{Susskind2004}
L.~Susskind and J.~Lindesay,
  \href{http://dx.doi.org/10.1142/9789812563095_0007}{``The stretched
  horizon,''} ,{\em An {{Introduction}} to {{Black Holes}}, {{Information}}
  and the {{String Theory Revolution}}},
World Scientific, (2004) 61--68.

\bibitem{Susskind:1994sm}
L.~Susskind and J.~Uglum, ``Black Hole Entropy in Canonical Quantum
  Gravity and Superstring Theory,''
  \href{http://dx.doi.org/10.1103/PhysRevD.50.2700}{{\em
  Phys. Rev. D} {\bfseries 50} (1994) 2700--2711},
  \href{http://arxiv.org/abs/hep-th/9401070}{{\ttfamily arXiv:hep-th/9401070}}.

\bibitem{Srednicki:1993im}
M.~Srednicki, ``Entropy and area,''
  \href{http://dx.doi.org/10.1103/physrevlett.71.666}{{\em Phys. Rev. Lett.}
  {\bfseries 71} (1993) 666--669}.

\bibitem{Hawking:2016sgy}
S.~W. Hawking, M.~J. Perry, and A.~Strominger, ``Superrotation {{Charge}} and
  {{Supertranslation Hair}} on {{Black Holes}},''
  \href{http://dx.doi.org/10.1007/JHEP05(2017)161}{{\em JHEP}
  {\bfseries 05} (2017) 161 },
  \href{http://arxiv.org/abs/1611.09175v2}{{\ttfamily arXiv:1611.09175v2}}.

\bibitem{He:2014laa}
T.~He, V.~Lysov, P.~Mitra, and A.~Strominger, ``{{BMS}} supertranslations and
  {{Weinberg}}'s soft graviton theorem,''
  \href{http://dx.doi.org/10.1007/JHEP05(2015)151}{{\em JHEP} {\bfseries 05} (2015) 151}.

\bibitem{Haco:2018ske}
S.~Haco, S.~W. Hawking, M.~J. Perry, and A.~Strominger, ``Black {{Hole
  Entropy}} and {{Soft Hair}},''
  \href{http://dx.doi.org/10.1007/JHEP12(2018)098}{{\em JHEP}
  {\bfseries 12} (2018) 098},
  \href{http://arxiv.org/abs/1810.01847}{{\ttfamily arXiv:1810.01847}}.

\bibitem{Hawking:2016msc}
S.~W. Hawking, M.~J. Perry, and A.~Strominger, ``Soft {{Hair}} on {{Black
  Holes}},'' \href{http://dx.doi.org/10.1103/PhysRevLett.116.231301}{{\em Phys.
  Rev. Lett.} {\bfseries 116} (2016) 231301},
  \href{http://arxiv.org/abs/1601.00921v1}{{\ttfamily arXiv:1601.00921v1}}.

\bibitem{Strominger2018}
A.~Strominger, \href{http://dx.doi.org/10.2307/j.ctvc777qv}{{\em Lectures on
  the Infrared Structure of Gravity and Gauge Theory}}.
\newblock Princeton University Press, (2018).

\bibitem{Cheng:2020vzw}
P.~Cheng and Y.~An, ``Soft black hole information paradox: {{Page}} curve from
  {{Maxwell}} soft hair of a black hole,''
  \href{http://dx.doi.org/10.1103/PhysRevD.103.126020}{{\em Phys. Rev. D} {\bfseries 103} (2020) 126020}.

\bibitem{Cheng:2022xyr}
P.~Cheng and P.~Mao, ``{Soft theorems in curved spacetime},''
  \href{http://dx.doi.org/10.1103/PhysRevD.106.L081702}{{\em Phys. Rev. D}
  {\bfseries 106} (2022) L081702},
  \href{http://arxiv.org/abs/2206.11564}{{\ttfamily arXiv:2206.11564
  [hep-th]}}.

\bibitem{Cheng:2022xgm}
P.~Cheng and P.~Mao, ``{Soft gluon theorems in curved spacetime},''
  \href{http://arxiv.org/abs/2211.00031}{{\ttfamily arXiv:2211.00031
  [hep-th]}}.

\bibitem{Barnich:2018zdg}
G.~Barnich, ``Black hole entropy from non-proper gauge degrees of freedom:
  {{II}}. {{The}} charged vacuum capacitor,''
  \href{http://dx.doi.org/10.1103/PhysRevD.99.026007}{{\em Phys. Rev. D} {\bfseries 99} (2018) 026007},
  \href{http://arxiv.org/abs/1806.00549v2}{{\ttfamily arXiv:1806.00549v2}}.

\bibitem{Barnich:2019qex}
G.~Barnich and M.~Bonte, ``Soft degrees of freedom, {{Gibbons-Hawking}}
  contribution and entropy from {{Casimir}} effect,''
  \href{http://arxiv.org/abs/1912.12698}{{\ttfamily arXiv:1912.12698}}.

\bibitem{Alessio:2020lpk}
F.~Alessio, G.~Barnich, and M.~Bonte, ``Gravitons in a {{Casimir}} box,''
\href{http://dx.doi.org/10.1007/JHEP02(2021)216}{{\em JHEP}
  {\bfseries 02} (2021) 216 },
  \href{http://arxiv.org/abs/2011.14432}{{\ttfamily arXiv:2011.14432}}.

\bibitem{Aggarwal:2022rrp}
A.~Aggarwal and G.~Barnich, ``{Phase transition of photons and gravitons in a
  {Casimir} box},'' \href{http://arxiv.org/abs/2205.03714}{{\ttfamily
  arXiv:2205.03714 [hep-th]}}.
\bibitem{Park:2022nkb}
I.~Y.~Park,
``Black hole entropy from non-Dirichlet sectors, and bounce solution,''
\href{http://arxiv.org/abs/2209.09074}{arXiv:2209.09074 [hep-th]]}.

\bibitem{carrollSpacetimeGeometry2019}
S.~M. Carroll, \href{http://dx.doi.org/10.1017/9781108770385}{{\em Spacetime
  and {{Geometry}}}}.
\newblock {Cambridge University Press}, (2019).

\end{thebibliography}

\end{document}